%% file: paper.tex
\documentclass[onecolumn]{article}                     
\usepackage{times}
\usepackage{enumerate}
\usepackage{bm,bbm,amsfonts,amssymb,amsmath}
\usepackage{graphicx}
\usepackage[T1]{fontenc}
\usepackage{gensymb}
\usepackage{hyperref}
\usepackage{url}
 \usepackage{mathptmx}      

\input{macro}

\newcommand\rpfc[1]{b_{#1}(\bpsi)\rangle}
\newcommand\rpf{\textbf{b}(\bpsi)\rangle}
\newcommand\rpfw{\rpfc{w}}
\newcommand\rpfwp{\rpfc{w'}}
\newcommand\lpf{\langle\textbf{b}(\bpsi)}
\newcommand\lpfw{\langle b_w(\bpsi)}
\newcommand\RPF{L(\bpsi)}
\newcommand\RPFww{L_{w' , w}(\bpsi)}
\newcommand\RPFnww{L^n_{w' , w}(\bpsi)}
\newcommand\spsi{s(\bpsi)}
\renewcommand\Con{\cyl{\omega}{0}{n-1}}
\renewcommand\pTo{{\pi^{(T)}}}
\newcommand\uom{\underline\omega}
\newcommand{\ands}{, }

\begin{document}

\title{Entropy-based parametric estimation of spike train statistics
}



\author{ J. C. Vasquez \thanks{ INRIA, 2004 Route des Lucioles, 06902 Sophia-Antipolis, France. \newline \indent   email: Juan-Carlos.Vasquez@sophia.inria.fr}, T. Vi\'eville $^\ast$, B. Cessac  $^\ast$ \thanks{ Laboratoire J. A. Dieudonn\'e,U.M.R. C.N.R.S. N\degree 6621, Universit\'e de Nice Sophia-Antipolis, France.}}


\maketitle

\begin{abstract}
We propose a generalisation
of the existing maximum entropy models used for spike trains statistics analysis, based
on the thermodynamic formalism from ergodic theory, and allowing one to  take into account
memory effects in dynamics. We propose a spectral method which provides
directly the ``free-energy'' density 
and the Kullback-Leibler divergence between the empirical statistics and the statistical model.
This method does not assume a specific Gibbs potential form. It does not require the assumption
of detailed balance and offers a control of finite-size sampling effects, inherent
to empirical statistics, by using large deviations results.  
A numerical validation of the method is proposed and the perspectives regarding spike-train code analysis are also discussed.\\
\newline
\noindent \textsl{\textbf{Keywords}: Spike train analysis \ands Higher-order correlation \ands Statistical Physics \ands Gibbs Distributions \ands Maximum Entropy}\\
\textsl{\textbf{PACS}: 05.10.-a  \ands 87.19.lo  \ands 87.19.lj}\\
\textsl{\textbf{MCS(2000)}: 37D35 \ands 37M25 \ands 37A30  }
\end{abstract}

\section{Introduction}
Processing and encoding of information in neuronal dynamics is a very active research
field \cite{rieke-warland-etal:97}, although still much of the role of neural assemblies and their internal interactions
remains unknown \cite{Pouget2008}. The simultaneously recording of the activity of
groups of neurons (up to several hundreds) over a dense configuration, supplies  a
critical database to unravel the role of specific neural assemblies. In complement of
descriptive statistics (e.g. by means of cross-correlograms or joint peri-stimulus time
histograms), somehow difficult to interpret for a large number of units (review in
\cite{brown-etal:04,kass-etal:05}), is the specific analysis of multi-units spike-patterns, as
found e.g. in \cite{Abeles88}. This approach develops algorithms to detect common
patterns in a data block, as well as performing combinatorial analysis to compute the expected
probability of different kind of patterns. The main difficulty with such type of approaches
is that they rely on a largely controversial assumption, Poissonian statistics (see
\cite{Pouzat2009a,Pouzat2009b,schneidman-etal:06}), which moreover, is a minimal
statistical model largely depending on the belief that firing rates are essentially the main
characteristic of spike trains.

A different approach has been proposed in \cite{schneidman-etal:06}.
They have shown that a model taking into account pairwise synchronizations between
neurons in a small assembly (10-40 retinal ganglion cells) describes most (90\%) of the
correlation structure and of the mutual information of the block activity, and performs
much better than a non-homogeneous Poissonian model. 
Analogous results were presented the same year in \cite{Shlens2006}. The
model used by both teams is based on a probability distribution known as the Gibbs
distribution of the Ising model which comes from statistical physics.
The parameters of this distribution relating, in neural data
analysis, to the firing rate of neurons and to their probability of
pairwise  synchronisation have to be determined from empirical data. Note that this approach has been
previously presented in neuroscience, but in a slightly different and more general
fashion, by \cite{Martignon2000,laskey-martignon:96,Martignon1995} (it was referred as ``log-linear models'').
The use of Ising model in neural decoding (especially of visual stimuli) has been largely exploited by
several other authors \cite{Cocco2009,Osborne2008,Shlens2009,tang-etal:08}. In particular, it is believed
by some of them \cite{Cocco2009} that  the pairwise coupling terms inferred from simultaneous
spikes corresponds,  in the model, to effective couplings between ganglion cells. In this spirit,
computing the parameters of the Ising model would provide an indirect access to ganglion cells connections. 
In addition, an increasing number of different theoretical and numerical developments of
this idea have recently appeared. In particular, in \cite{tkacik-schneidman-etal:06}, the
authors propose a modified learning scheme and thanks to concepts taken from physics,
such as heat capacity, explore new insights like the distribution of the underlying density of states; additionally in \cite{roudi-tyrcha-etal:09,roudi-aurell-etal:09} authors study and  compare several approximate, but faster, estimation methods for learning the couplings and apply them on experimental and synthetic data drawing several results for this type of modeling.

However, it might be questionable whether more general
form of Gibbs distributions (e.g. involving $n$-uplets of neurons) could improve the estimation
and account for deviations to Ising-model (\cite{Shlens2009,tkacik-schneidman-etal:06}) and provide a better understanding of the neural code from the point of view of the maximal entropy principle \cite{jaynes:57}.
As a matter of fact, back to 1995, \cite{Martignon1995} already considered multi-unit synchronizations
and proposed several tests to understand the statistical significance of those synchronizations
and the real meaning of their corresponding value in the energy expansion.
A few years later, \cite{Martignon2000} generalized this approach
to arbitrary spatio-temporal spike patterns and compared this method to
other existing estimators of high-order correlations or Bayesian approaches. They also introduced a
method comparison based on the Neyman-Pearson hypothesis test paradigm. Though
the numerical implementation they have used for their approach presented strong
limitations, they have applied this methods successfully to experimental data from
multi-units recordings in the pre-frontal cortex, the visual cortex of behaving monkeys,
and the somato-sensory cortex of anesthetized rats.
Several papers have pointed out the importance of temporal patterns of activity at the network level \cite{lindsey-morris-etal:97,villa-tetko-etal:99,segev-baruchi-etal:04}, and recently \cite{tang-etal:08} have shown the insufficiency 
of Ising model to predict the temporal statistics of the neural activity. 
As a consequence, a few authors (we know only one reference, \cite{marre-boustani-etal:09})
have attempted to define time-dependent Gibbs distributions on the base of a Markovian approach
(1-step time pairwise correlations)
and convincingly showed a clear increase in the accuracy of the spike train statistics characterization.
Namely, this model produces a lower Jensen-Shannon Divergence, when analyzing
raster data generated by a Glauber spin-glass model, but also \textit{in vivo} multineuron
data from cat parietal cortex in different sleep states. \\ 

To summarize, the main advantages of all these 'Ising-like' approaches are:

\bit
\item{(i)} to be based on a widely used principle,
the maximal entropy principle \cite{jaynes:57} to determine statistics from the empirical knowledge
of \textit{(ad hoc)} observables; 
\item{(ii)} to propose statistical models having
a close analogy with Gibbs distributions of magnetic systems, hence disposing of several deep theoretical results and
numerical methods (Monte-Carlo methods, Mean-Field approximations, series
expansions),  resulting in a fast analysis of experimental data from
large number of neurons.
\eit

\smallskip

However,  as we argue in this paper, this approaches presents also, in its current state, fundamental weaknesses:
\bit
\item{(i)} The maximal entropy principle leads, in its classical formulation,
 to a parametric form, corresponding to chosing a 
finite set of \textit{ad hoc} constraints, which only provides an approximation of the real
statistics, while the distance (say measured by the Kullback-Leibler divergence) between
the model and the hidden distribution can be quite large  \cite{csiszar:84}. 
Moreover, when considering time dependent correlations,
 this procedure leads to Gibbs potential which requires a proper renormalisation
in order to be related to a Markov chain (see section \ref{Snorm}).

\item{(ii)} The  Gibbs distributions considered by these approaches, with
the naive form ``$\frac{1}{Z}e^{-\beta H}$'', where $Z$ is a constant (while
it depends on boundary terms in the general case) have a limited degree of application;
in particular they do not extend easily to time dependent sequences with long memory,
as spikes train emitted from neural networks might well be.
 Especially, considering already one time step Markov processes
leads to substantial complications a shown in  \cite{marre-boustani-etal:09}.
The ``partition function'' is not a constant
(see eq. (1) of paper \cite{marre-boustani-etal:09}) and needs to be approximated
(eq. (4) of the same paper) using the (unproved) assumption of detailed balance, which is moeover a sufficient
but non necessary condition for the existence of an equilibrium state, and may hardly generalize
to more elaborated models. 
\item{(iii)} It does not allow to treat in a straightforward way the time-evolution of
the Gibbs distribution (e.g. induced by mechanisms such as synaptic plasticity).
\eit 

\smallskip

However, more general forms of Gibbs distributions have been introduced since long
\cite{sinai:72,ruelle:78,bowen:98}, in a theory called ``thermodynamic formalism'' introduced
 in the realm of dynamical systems and ergodic
theory, allowing to treat infinite time sequences of processes with long (and even infinite
 \cite{maillard:07}) memory. In this paper, we use the thermodynamic formalism to propose a generalisation
of the existing models used for spike trains statistics analysis which results in a more powerful framework that overcomes some of
the weaknesses mentioned above. Our results
are grounded on well established theoretical basements (see e.g. \cite{keller:98})
completed by recent results of the authors dealing with collective spike trains statistics
in neural \textit{networks} models \cite{cessac-rostro-etal:09,cessac:10}.
The theoretical framework of our approach is presented in the section \ref{STheory}.
We propose a global approach to spike train analysis,
going beyond the usual approaches essentially because it allows us to  take into account
(long term) \textit{memory effects} in dynamics (sections \ref{SGen},\ref{SMarkov}). As a matter of fact
we deal with models considering \textit{spatio-temporal}
and time-\textit{causal} structure of spike trains emitted by neural networks
together with the fact that some spike sequences (or ``words'') might be forbidden by dynamics, introducing
the notion of \textit{grammar}. We propose a spectral method which provides
directly the ``free energy density'' 
and the Kullback-Leibler divergence between the empirical statistics and the statistical model (section
\ref{SStat}).
This method does not assume a specific potential form and allows us to handle correctly
non-normalized potentials. It does not require the assumption
of detailed balance (necessary to apply Markov Chain Monte-Carlo (MCMC) methods) and offers a control of finite-size sampling effects, inherent
to empirical statistics, by using large deviations results (Section \ref{Semp}).  
The price to pay is to introduce a somewhat heavy, but necessary, mathematical formalism.
In several places we make connections with existing methods to clarify these concepts.

These theoretical basements allows us to propose, in section \ref{Methods},
 a numerical method  to
parametrically estimate, and possibly compare, models for the statistics of simulated
multi-cell-spike trains. Our method
is not limited to firing rates models, pairwise synchronizations as \cite{schneidman-etal:06,Shlens2006,Shlens2009} or
$1$-step time pairwise correlations models as \cite{marre-boustani-etal:09}, but
deals with general form of Gibbs distributions, with parametric potentials corresponding
to a spike $n$-uplets expansion, with
multi-units and multi-times terms. The method is exact (in the sense that is does not
involve heuristic minimization techniques).
Moreover, we perform fast and reliable estimate of quantities such as the Kullback-
Leibler divergence allowing a comparison between different models, as well as the computation of standard statistical
indicators, and a further analysis about convergence rate of the empirical
estimation and large deviations.

 In section \ref{Results}
we perform a large battery of tests allowing us to experimentally validate the method. First, we analyse the numerical precision of parameter estimation.
Second, we generate synthetic data with a given statistics, and compare the estimation obtained  using these data for
several models. Moreover, we simulate a neural network and propose the estimation of the underlying Gibbs distribution  parameters whose analytic form is known \cite{cessac:10}. We also perform the estimation for several models using data obtained from a simulated neural network  with stationary dynamics after Spike-Time dependent synaptic plasticity. Finally, we show results on the parameters estimation from synthetic data generated by a non-stationary statistical model.

\section{Spike trains statistics from a theoretical perspective.}\label{STheory}
\subsection{General context}\label{SGen}

\subsubsection{Neural network dynamics.}

We consider the evolution of a network of $N$ neurons, described by a 
dynamical model, that is,  either a deterministic dynamical system or
a stochastic dynamical system (usually governed by both a deterministic evolution map and additive noise).
We assume that there is a minimal time scale $\delta$, set to $\delta=1$ without loss of generality, 
at which dynamics can be time-discretized. Typically, this can be the minimal resolution of the spike
time, constrained  by biophysics and by measurements methods
 (see \cite{cessac:07} for a discussion on time discretisation in spiking neural networks).
The typical neuron models we think of are punctual conductance based generalized Integrate-and-Fire (IF) models with exponential synapses (gIF) \cite{cessac-vieville:08}.
Actually, the formalism developed here has been rigorously
funded in  \cite{cessac:10} for Leaky-Integrate-and-Fire (LIF) models with noise.
We further assume the network parameters (synaptic weights, currents, etc..) to be fixed in this context 
(see \cite{cessac-vieville:08} for a discussion).
This means that we assume observing a period of time where the system parameters are essentially constants.
In other words, we focus here on \textit{stationary} dynamics. This restriction is further discussed
in section \ref{SNonStat}.

We are interested in situations where neurons dynamics, and especially spikes occurrences, do not show 
any regularity or exact reproducibility and require a statistical treatment. This
is obviously the case for stochastic evolutions but this also happens in the deterministic case, 
whenever dynamics exhibits initial conditions sensitivity.
This leads us to the choice of the statistical formalism proposed here, called the ``thermodynamic
formalism\footnote{This terminology has been introduced by Sinai \cite{sinai:72}, Ruelle \cite{ruelle:78}
and Bowen \cite{bowen:98} because of its analogy with statistical physics. 
But it does not relies on the principles of thermodynamics. Especially, the maximization of statistical entropy,
discussed below, does not requires the invocation of the second principle of thermodynamics.} ''
(see \cite{cessac-rostro-etal:09} for an extended discussion).

\subsubsection{Dynamics and raster plots.}\label{SDynDef}
Each neuron of index $i=0 \dots N-1$ is characterized by
its state, $X_i$, which  belongs to some (bounded) set $\cI \in \bbbr^M$.
  $M$ is the number of variables characterizing
the state of one neuron (we assume that all neurons are described by the same number of variables).
A typical example is  $M=1$ where $X_i=V_i$ is the membrane potential
of neuron $i$ and $\cI=[\Vm,\VM]$ but the present formalism affords
extensions to such additional characteristics as activation variables (e.g. for the Hodgkin-Huxley model \cite{hodgkin-huxley:52} $M=4$). 
The variable  $X=\left[X_i\right]_{i=0}^{N-1}$  represents the  state of a network
of $N$ neurons. 
Without loss of generality, we assume that all neurons have the same properties so
that $X \in \cM=\cI^N$, where $\cM$ is the  phase space where dynamics occurs. The evolution of the network over an infinite time is characterized
by a \textit{trajectory} $\tilde{X} \deq \left\{X(t)\right\}_{t=-\infty}^{+\infty}$.


One  can associate to each
neuron $i$ a variable $\omega_i(t)=1$ if neuron $i$ fires between $[t,t+1[$ and  $\omega_i(t)=0$
otherwise.  A ``spiking pattern'' is a vector  $\omega(t) \deq \left[\omega_i(t) \right]_{i=0}^N-1$
which tells us which neurons are firing at time $t$. 
%
%
In this setting, a ``raster plot'' 
 is a sequence $\tom \deq \left\{\omega(t)\right\}_{t=-\infty}^{+\infty}$,
of spiking patterns. 
Finally a {\em spike block} is a finite set of spiking pattern, written:
$$\bloc{t_1}{t_2} = \left\{\omega(t)\right\}_{\{t_1 \leq t \leq t_2\}},$$
where spike times have been prescribed between the times $t_1$ to $t_2$.

To each trajectory $\tilde{X} = \left\{X(t)\right\}_{t=-\infty}^{+\infty}$
 is associated a raster plot $\tom=\left\{\omega(t)\right\}_{t=-\infty}^{+\infty}$. This is the sequence of spiking patterns
displayed by the neural network when it follows the trajectory $\tilde{X}$.
We write $\tilde{X} \to \tom$.
On the other way round, we say that an infinite sequence $\tom=\left\{\omega(t)\right\}_{t=-\infty}^{+\infty}$
is \textit{an admissible raster plot} if dynamics allows a trajectory
$\tilde{X}$ such that   
$\tilde{X} \to \tom$. We call $\Sigma$ the set of admissible raster plots.
The dynamics of the neurons state induces therefore a dynamics on the set of admissible raster plots,
represented by the 
 \textit{left shift}, $\sigma$, such that $\sigma \tom = \tom' \Leftrightarrow
\omega'(t)=\omega(t+1), \forall t \geq 0$ 
%
.
Thus,  in some sense,
raster plots provide a code for the trajectories $\tilde{X}$. 
Note that the correspondence may not be one-to-one \cite{cessac:08}.

Though dynamics produces many possible raster plots, it is important
 to remark that it is not able to produce {\em any} possible sequence of spiking patterns.
This depends on the system properties (e.g., refractoriness forbids raster plots with spike interval below $1ms$) 
and  parameters (e.g., after synaptic weight adaptation the dynamics often appears more constrained).
For example,
inhibition may prevent a neuron to fire whenever a group of pre-synaptic neurons
has fired before. 
There are therefore \textit{allowed} and \textit{forbidden} sequences, constrained by dynamics.
This corresponds to the following \textit{crucial} property, often neglected
in entropy estimations of spike trains \cite{rieke-warland-etal:97}. The set
of admissible raster plot $\Sigma$ is \textit{not the set of all possible
raster plots}. Indeed,  considering spike blocks of size $n$ there 
are $2^{Nn}$ possible spike blocks but quite a bit less 
\textit{admissible}
raster plots (the exponential rate of growths in the number of
admissible raster plots is given by the topological entropy which
is an upper bound for the Kolmogorov-Sinai entropy defined in eq. (\ref{hKS}), footnote \ref{notehKS}).
\subsubsection{Transition probabilities.}\label{STrans}

Typically, the network dynamics and the related spikes fluctuate in an unpredictable manner.
The spike response itself is not sought as a deterministic response in this context, but as a conditional probability \cite{rieke-warland-etal:97}. 
``Reading out the code'' consists of inferring such probability.
Especially, the probability that a neuron emits a spike at some time $t$
depends on the history of the neural network.
However,  it is impossible to know explicitely its form in the general case
 since it depends on the past evolution of all  variables determining 
the neural network state $\X$. A possible simplification is to consider that this probability depends
\textit{only} on the spikes emitted in the past by the network. In this way,
we are seeking a family of transition probabilities of the form
$P\left[\omega(t) \, | \, \bloc{-\infty}{t-1} \right]$ from which all spike trains statistical properties
can be deduced.  These transition probabilities are called \textit{conditional intensity}
in \cite{johnson-swami:83,brillinger:88,chornoboy-schramm-etal:88,kass-ventura:01,truccolo-eden-etal:05,okatan-wilson-etal:05,truccolo-donoghue:07,pouzat-chaffiol:09} and they are essential to determine completely the spike trains statistics.
The price to pay is that we have to consider processes with memory (which is not so shocking
when dealing with neural networks).

These transition  probabilities are unknown for most models
but an explicit computation can be rigorously achieved in the case of a discrete time Leaky Integrate-and-Fire (LIF) neural networks with noise, in the stationary case (e.g. time independent stimulus) (see eq. (\ref{phiBMS}) and \cite{cessac:10}).  Stationarity means here that the transition probability does not depend explicitely on $t$ so that one can focus on transition probabilities of the form $P\left[\omega(0) \, | \, \bloc{-\infty}{-1} \right]$ and infer the probability of any spike block by the classical Chapman-Kolmogorov equation \cite{gikhman-skorokhod:79}. 
To our knowledge this is
the only example where the complete spike trains statistics can be rigorously and analytically computed. Note that the transition probability depends on a \textit{unbounded} past in the LIF model. 
Indeed, the state of a neuron is reset whenever it fires, so the probability of a given spiking pattern at time $0$ depends on the past up to a time when each neuron has fired at least once. However, this time cannot be bounded (though the probability that it is larger than some $\tau$ decays exponentially fast with $\tau$) \cite{cessac:10}.

\subsubsection{Gibbs distribution.}\label{SGibbs}

As far as the present paper is concerned, the main result in \cite{cessac:10} states that
some neural networks models \textit{do have Gibbs distributions}, though of a quite more complex
form than currently used in the literature. More precisely it is rigorously proved
in \cite{cessac:10} that in
discrete-time LIF models\footnote{Without restriction on the synaptic weights except that they are finite.} with noise
 the statistics of spike trains is characterized by a \textit{Gibbs distribution}  
which is also an \textit{equilibrium state}, where the potential can be \textit{explicitely computed}, but 
\textit{has infinite range}.
 
Let us be more explicit.   Since we are using the terms ``Gibbs distribution'' and ``equilibrium state'' in a 
more general sense than the definition used in the neuroscience community for spike train statistics analysis, we give here the definition of these two terms. In several places in the paper we show the link
between this formalism and the usual form, and explain why we need to use the present formalism
for spike train analysis.
The main difference is that we consider  
probability distributions on a set of spatio-temporal sequences where the ``space'' is the network, 
and where \textit{time is infinite} so that the spike train probability distributions is defined 
on infinite time sequences \footnote{This corresponds to the ``thermodynamic limit'' in statistical physics
but in our case thermodynamic limit means ``time tends to infinity'' instead of 
``dimension of the system tends to infinity''. As a matter of fact the number of neurons, $N$, is fixed and
finite in the whole paper.}. This is the natural context when considering transition probabilities
as introduced in the previous section.  The price to pay is a more complex formulation than
the classical $\frac{1}{Z}\exp(-\beta H)$, but the reward is having a formalism allowing us to handle
spike trains statistics including memory terms, and an explicit way 
to compute the free energy density and the Kullback-Leibler divergence between the empirical statistics
and a statistical model, as developped in the rest of the paper.

A probability distribution $\mu_\phi$ 
on the set of \textit{infinite} spike sequences $\Sigma$ (raster plots) is a Gibbs distribution if there exists
a function\footnote{Some regularity conditions, associated with a sufficiently fast decay
of the potential at infinity, are also required, that we do not state here \cite{keller:98}. \label{Footpot}} $\bphi :  \Sigma \to \setR$, called a 
\textit{potential}, such that the probability of a spike block
$\bloc{t}{t+n}$, for any $-\infty < t < +\infty$, and $n >0$, obeys:
\beq\label{defGibbs}
c_1 \leq \frac{\mu_\phi\left(\bloc{t}{t+n}\right)}
{\exp \left[ -(n+1) P(\bphi)+\sum_{k=t}^{t+n}\bphi(\sigma^k\omega) \right]}  \leq c_2,
\eeq
where $P(\bphi),c_1,c_2$ are some constants with $0 < c_1 \leq 1 \leq c_2$.
Recall that $\sigma$ is the shift on rasters defined in section \ref{SDynDef}.
Basically, this expresses that, as $n$ becomes large, $\mu_\phi\left(\bloc{t}{t+n}\right)$ behaves\footnote{In the sense of (\ref{defGibbs}).Thus,''behaves like'' does not mean ``is equal to''.} like 
$\frac{\exp\left[\sum_{k=t}^{t+n}\bphi(\sigma^k\omega) \right]}{ \exp\left[ (n+1) P(\bphi)\right]}$.

An \textit{equilibrium state} is a probability distribution $\mu_\phi$ which satisfies the following variational principle:
\beq \label{VarPrinc}
P(\bphi) \deq h(\mu_\phi)+\mu_\phi(\bphi)=\sup_{\mu \in m^{(inv)}(\Sigma) } h(\mu)+\mu(\bphi),
\eeq
where $m^{(inv)}(\Sigma)$ is the set of invariant probability measures on $\Sigma$,
$h(\mu)$ is the entropy\footnote{\label{notehKS} 
The Kolmogorov-Sinai entropy or entropy rate of a probability $\mu$ is: 

\beq\label{hKS}
h\left[\mu\right]=\lim_{n \to + \infty} \frac{h^{(n)}\left[\mu\right]}{n},
\eeq

\nid where
\beq\label{hKS2}
h^{(n)}\left[\mu\right]=-\sum_{\omega \in \Spn} \mu\left(\cyl{\omega}{0}{n-1}\right)\log\mu\left(\cyl{\omega}{0}{n-1}\right),
\eeq
$\Spn$ being the set of admissible sequences of length $n$.
This quantity provides the exponential rate of growth of admissible blocks having a positive probability under $\mu$, as $n$ growths.
It is positive for chaotic system and it is zero for periodic systems.} of the probability $\mu$,
and $\mu(\phi) \deq \int \phi d\mu$ is the average of $\phi$ with respect to the probability $\mu$.  
Note that the notion of Gibbs distribution and equilibrium state are not equivalent in general \cite{keller:98}, but
in the present context, they are\cite{cessac:10}.

The term $P(\bphi)$, called the \textit{topological pressure} in this context, is the formal
analog of a thermodynamic potential (free energy density). It is a generating function for the cumulants
of $\bphi$ (see section \ref{Sexamples} for explicit examples). \\

\sssu{Gibbs potential.}
In the case of discrete-time LIF models the potential $\phi$ is the log of the
probability  transition $P\left[\omega(t) \, | \, \bloc{-\infty}{t-1} \right]$ \cite{cessac:10}. We believe that
this statement extends to more general situations: if a spike train is characterized by a Gibbs distribution
then a natural candidate for the Gibbs potential is the log of the conditional intensity.
Let us insist on this result. Beyond the mathematical intrincacies grounding this statement,
this choice is natural because it provides a (time) \textit{causal} potential with \textit{memory}.
 As a consequence, the statistics
of spikes at a given time are causaly related to the past spikes.
This corresponds to potential having a \textit{range} that can be large.  A potential
has range $R$ if $\phi(\bloc{-\infty}{0})=\phi(\bloc{-(R-1)}{0})$. In terms of the transition probability,
this corresponds to a system with a memory depth $R-1$ (the probability that a neuron spike
at time $t$ depends on the spikes emitted by the network, up to time $t-(R-1)$ back in the past\footnote{Hence range $1$ or equivalently memory depth $0$ means time independent events.}). Unfortunately
even if the simplest known example of neural network model, the LIF,
the range is (mathematically) infinite\footnote{Though the variation of $\phi$ decays exponentially fast
ensuring the existence of a thermodynamic limit.}. Is the situation simpler
for more complex neural networks models, for real neural networks ?
Fortunately, finite range approximations can be proposed, with a good control
on the degree of approximation, as we now develop.

\subsection{Markov approximations.}\label{SMarkov}

In the sequel, we make the assumption that the spike trains statistics of the system that we are observing
is described by a Gibbs distribution whose potential has
to be determined from empirical data.

\subsubsection{Range-$R$ potential.} 
It is always possible to propose Markov approximations of $\bphi$, even in the case where the Gibbs potential depends on spike sequences with unbounded length. 
This is the strategy that we now develop.
We approximate the exact transition probability by a transition probability with finite memory
of depth $R-1$, $P\left[\omega(0) \, | \, \bloc{-(R-1)}{-1} \right]$. In this context, 
as shown in \cite{cessac:10},  the exact Gibbs
potential can be approximated\footnote{In the case of LIF models the Kullback-Leibler divergence
between the exact Gibbs distribution and its approximation by the potential (\ref{potential_expansion}) decays
exponentially fast with $R$.} by a  \textit{range-$R$ potential}  with a parametric form:
\beq\label{potential_expansion}
\bpsi(\omega)=
\sum_{l=1}^{R}
\sum_{
\tiny{
(i_1,t_1), \dots,(i_l,t_l) \in \cP(N,R),\\
}
}
\lambda^{(l)}_{i_1,n_{i_1},\dots,i_l,n_{i_l}} 
\omega_{i_1}(n_{i_1}) \dots\omega_{i_l}(n_{i_l}).
\eeq
This form is nothing but a Taylor expansion of  $\log(P\left[\omega(0) \, | \, \bloc{-(R-1)}{-1} \right])$,
where one collects all terms of form $\omega^{k_1}_{i_1}(n_{i_1}) \dots\omega^{k_l}_{i_l}(n_{i_l})$,
for integer $k_1, \dots k_l$'s, using that $\omega^{k}_i(n)=\omega_i(n)$, for any
$k > 0$ and any $i,n$.
  Here $\cP(N,R)$ is
the set of non repeated pairs of integers $(i,n)$ with
$i \in \left\{0,\dots,N-1 \right\}$ and 
$n  \in \left\{0\dots,R-1 \right\}$.

Such form of potential is a linear combination of \textit{monomials}.
An \textit{order-$n$ monomial} is
a product $\omega_{i_1}(t_1) \dots \omega_{i_n}(t_n)$,
where $0 \leq i_1 \leq i_2 \leq \dots \leq i_n \leq N-1$,
$0 \leq t_1 \leq t_2 \leq \dots \leq t_n < \infty$
and such that there is no repeated pair $(i_k,t_k)$, $k=1 \dots n$.
The monomial $\omega_{i_1}(t_1) \dots \omega_{i_n}(t_n)$
takes values in $\left\{0,1\right\}$ and is $1$ if and  only
if each neuron $i_l$ fires at time $t_l$, $l=1 \dots n$.
On phenomenological grounds the monomial $\omega_{i_1}(t_1) \dots \omega_{i_n}(t_n)$ corresponds
to a spike $n$-uplet $(i_1,t_1), \dots, (i_n,t_n)$
(neuron $i_1$ fires at time $t_1$, neuron $i_2$ at time $t_2$, etc ...). 

\sssu{Further approximations.}
The potential (\ref{potential_expansion}) remains quite cumbersome
since the number of terms in (\ref{psi}) explodes combinatorially as $N,R$ growth. 
Equivalently, in terms of the classical Jaynes approach where the Gibbs distribution
is obtained via the maximisation of statistical entropy under constraints (see section
\ref{SJaynes}), one has to fix a number of constraints that growths
exponentially fast with $N,R$.
As a consequence,
one is typically lead to consider parametric forms where  monomials
have been removed (or, sometimes, added) in the expansion.  This constitutes a coarser approximation to
the exact potential, but more tractable from the numerical or empirical point of view. To alleviate notations
we write, in the rest of paper, the parametric potential in the
form:
\beq\label{psi}
\bpsi=\sum_{l=1}^L \lambda_l \phi_l,
\eeq
where $\phi_l$'s are monomials. The choice of the parametric form defines
what we call a ``statistical model'', namely a Gibbs distribution,   denoted $\mpg$ in the sequel, for the potential
(\ref{psi}). The question is ``how far is this distribution from the true statistics'' ?

\sssu{Examples of range-$R$ potentials}\label{Sexamples}

\textbf{Bernoulli potentials} The easiest example of potential are range-$1$ potentials (memoryless)
where $\bpsi(\tom)=\sum_{i=0}^{N-1}\lambda_i \omega_i(0)$. The corresponding Gibbs distribution 
provides a statistical model where neurons are independent.\\

\nid \textbf{``Ising'' like potentials.} This type of potential has been introduced by Schneidman and collaborators in \cite{schneidman-berry-etal:06}. 
It reads, in our notations,
\beq\label{Ising}
\bpsi(\tom)=\sum_{i=0}^{N-1} \lambda_i \omega_i(0)+
\sum_{i=0}^{N-1}\sum_{j=0}^{i-1} \lambda_{ij}  \omega_i(0) \omega_j(0).
\eeq
The corresponding Gibbs distribution 
provides a  statistical model  where synchronous pairwise correlations between neurons
are taken into account, but neither higher order spatial correlations nor other time correlations
are taken into account. As a consequence, the corresponding ``Markov chain'' is memoryless.\\

\nid \textbf{Pairwise Time-Dependent-$k$  potentials with rates (RPTD-$k$).}

An easy generalization of (\ref{Ising}) is:
\beq \label{RPTD-k}
\bpsi(\tom)=\sum_{i=0}^{N-1} \lambda_i \omega_i(0) +
\sum_{i=0}^{N-1}\sum_{j=0}^{i-1} \sum_{\tau=-k}^{k} \lambda_{ij\tau}  \omega_i(0) \omega_j(\tau),
\eeq
called \textit{Pairwise Time-Dependent $k$ (RPTD-$k$)  with Rates} potentials in the sequel.\\

\nid \textbf{Pairwise Time-Dependent $k$ (PTD-$k$) potentials.}

A variation of (\ref{RPTD-k}) is to avoid the explicit constraints associated to firing rates :
\beq \label{PTD-k}
\sum_{i=0}^{N-1}\sum_{j=0}^{i-1} \sum_{\tau=-k}^{k} \lambda_{ij\tau}  \omega_i(0) \omega_j(\tau),
\eeq
called \textbf{Pairwise Time-Dependent $k$ (PTD-$k$) } potentials in the sequel.

\sssu{Encoding spike blocks} 
To each spike block of length $R$, $\bloc{k}{k+R-1}$, $k \in \setZ$,  one can associate
an integer:
\beq \label{omtow}
w_k
=\sum_{t=k}^{k+R-1} \sum_{i=0}^{N-1} 2^{i+Nt}\omega_i(t).
\eeq
One has  $2^{NR}$ such possible blocks (though some of them can be forbidden by dynamics).

We use the following notation:
\beq \label{somtosw}
\sigma w_k = \sum_{t=k+1}^{k+R} \sum_{i=0}^{N-1}2^{i+Nt}\omega_i(t),
\eeq
so that, $w_k$ represents the block $\bloc{k}{k+R-1}$ 
and $\sigma w_k=w_{k+1}$  represents the block $\bloc{k+1}{k+R}$.
In this setting a range-$R$ potential is therefore a vector in the space $\cH \deq \bbbr^{2^{NR}}$
with components $\psi_w \deq \bpsi(\omega)$. 
This amounts to recoding spiking sequences as sequences of spike blocks of length $R$, associated with
words $w_k$, taking into account
the memory depth of the Markov chain.

\ssu{Determining the statistical properties of a Gibbs distribution.}\label{SStat}

We now introduce the thermodynamic formalism allowing us to compute numerically
the main statistical properties of a Gibbs distribution. This approach
is different from a classical approach in statistical physics where
one tries to compute the partition function. The present
approach gives directly the topological pressure (corresponding to the free energy
density in the thermodynamic limit) from which the statistical properties can be inferred.

\subsubsection{The Ruelle-Perron-Frobenius operator.}\label{SRPF}

Once the parametric form of the potential is given, the  statistical properties of the Gibbs distribution
are obtained by the  Ruelle-Perron-Frobenius (RPF) operator  introduced by Ruelle in \cite{ruelle:68}.
In the present case  this is a positive $2^{NR} \times 2^{NR}$ matrix, $\RPF$ , with entries 
\beq\label{defRPF}
\RPFww=
e^{\bpsi_{w'}} G_{w',w} \quad,
\eeq
(while it acts 
on functional spaces in the infinite range case). 

The matrix $G$ is called the \textit{grammar}.
It encodes  in its definition the \textit{essential fact} that the underlying dynamics
is not able to produce all possible raster plots:

\beq\label{grammar}
G_{w',w}=
\left\{
\baR{lll}
1, \quad \, \mbox{if the transition} \, w' \to w \, \mbox{is admissible};\\
0, \quad \mbox{otherwise}.
\eaR
 \right.
\eeq

Since we are considering blocks of the form\footnote{Since dynamics is assumed stationnary 
the result actually does not depend on $k$.} $w' \sim \bloc{k}{k+R-1} = \bom(k) \dots \bom(k+R-1)$
and $w \sim \bloc{k+1}{k-R} = \bom(k+1) \dots \bom(k-R)$, the transition $w' \to w$ is legal
if\footnote{Additional transitions are usually forbidden by dynamics. As a consequence,
those transitions have a zero probability of occurence and they can be detected on empirical sequences
(see section \ref{SMainProp}).} $w'$ and $w$ have the spiking patterns $\bom(k+1) \dots \bom(k+R-1)$ in common. Thus,  while there
are $2^{NR}$ blocks for a network of $N$ neurons, the matrix $G$ has
at most $2^{N}$ non zero entries on each line.
As a consequence $\RPF$ is  \textit{sparse}.

Note also that all non zeroes entries $\RPFww$ on a given line are equal.
This degeneracy comes from our choice to represent $\bpsi$ as a vector in $\cH$
which is the easiest for numerical purposes. This has consequences
discussed in the section \ref{Snorm}.

\sssu{The Ruelle-Perron-Frobenius theorem}\label{SRPFTh}
In the present paper we make the assumption 
that the underlying (and hidden) dynamics
is  such that  the $\RPF$ matrix is primitive,
i.e. $\exists n>0$, s.t. $\forall w,w'$ $\RPFnww>0$.
This assumption holds
for Integrate-and-Fire models with noise 
and is likely to hold for more general neural networks
models where noise renders dynamics ergodic and mixing
\cite{cessac:10}. Note, on the opposite, that if this assumption
is not fulfilled there are little chances to characterize
spike trains statistics with a (unique) Gibbs distribution.

Then, $\RPF$
obeys the Perron-Frobenius theorem\footnote{This 
theorem has been generalized by Ruelle to
infinite range potentials under some regularity conditions
 \cite{ruelle:69,ruelle:78}.}:

\bth  $\RPF$ has a unique maximal and strictly positive eigenvalue $\spsi=e^{P(\bpsi)}$
 associated with a right eigenvector $\rpf$  
and a left eigenvector $\lpf$, with positive and bounded entries, such that
$\RPF \rpf=\spsi\rpf$, $\lpf \RPF =\spsi\lpf$. Those vectors can be chosen such
that 
$\lpf.\rpf=1$  where $.$ is the scalar product in $\cH$.
The remaining part of the spectrum is located in a disk in the complex plane, of radius strictly lower than $\spsi$.
As a consequence, for all $\bv$ in $\cH$, 

\beq \label{ConvVect}
\frac{1}{\spsi^n} L^n(\bpsi) \bv \to \rpf \lpf \, . \, \bv,
\eeq

\nid as $n \to \infty$.

The Gibbs-probability of a spike block $w$ of length $R$ is 

\beq\label{invmeascomp}
\mu_{\bpsi}(w)=\rpfw \lpfw,
\eeq

\nid where $ \rpfw$ is the $w$-th component of $\rpf$.
\enth

As a consequence, the assumption of primitivity guarantees the existence and uniqueness of a Gibbs
distribution. Note that it is more general than the detailed balance assumption.

\subsubsection{Computing averages of monomials}  
Since $\mpg[\phi_l]=\sum_{w} \mpg[w] \phi_l(w)$ one obtains
using (\ref{invmeascomp}):
\beq\label{avobservable}
\mpg[\phi_l]=\sum_{w \in \cH} \rpfw \phi_l(w)\lpfw.
\eeq
This provides a fast way to compute  $\mpg[\phi_l]$.

\subsubsection{The topological pressure.} \label{Spressure}

The RPF theorem gives a direct access to the
topological pressure $\pres$ which is the logarithm of the leading
eigenvalue $\spsi$, easily obtained by a power method (see eq. (\ref{ConvVect})).
In the case of range-$R$ potentials (\ref{psi}) where the topological pressure $\pres$ 
becomes a function of the parameters $\bl=\left(\lambda_l\right)_{l=1}^L$,
 we write $P(\bl)$. One can show that the topological pressure 
is the generating function for the cumulants of the monomials $\phi_l$:

\beq\label{Gener}
\frac{\partial  P(\bl)}{\partial \lambda_l}=\mpg[\phi_l].
\eeq

Higher order cumulants 
are obtained likewise by successive derivations. 
Especially, second order moments related to the central limit
theorem obeyed by Gibbs distributions \cite{bowen:98,keller:98}
are obtained by second order derivatives.
As a consequence of this last property, the topological pressure's Hessian is positive
and  the topological pressure is \textit{convex} with respect to $\bl$.

\subsubsection{Entropy} Since $\mpg$ is a Gibbs distribution, for the potential
$\bpsi$, therefore, an exact expression for the Kolmogorov-Sinai entropy
(\ref{hKS}) can be readily obtained:

\beq\label{CalchKS}
h\left[\mpg\right]=P(\bl)-\sum_{l} \lambda_l \mpg\left[\phi_l\right].
\eeq

\sssu{Normalisation}\label{Snorm}

When switching from the potential (\ref{potential_expansion}), which
is the polynomial expansion of the log of the conditional intensity,
to a simplified parametric form (\ref{psi}), one introduces several
biases. First, one may add terms which are not in the original potential.
Second,  (\ref{potential_expansion}) must satisfy a constraint
corresponding to the fact that $\bpsi(\omega)$ is the $\log$ of a probability. Such a potential
is called \textit{normalized}. Its main characteristics are (i) the topological
pressure is zero; (ii)  the right eigenvector $\rpf$ has all
its components equal. The reason is simple: when the potential is
the log of a transition probability the RPF operator satisfies
$\sum_{w \in \cH} L_{w'w}=\sum_{\omega(0)\in \left\{0,1 \right\}^N} P\left[\omega(0) \, | \, \bloc{-(R-1)}{-1} \right]=1,
 \forall w' \in  \cH$, where $w'$ corresponds e.g. to 
$\bloc{-R}{-1}$, and $w$ to $\bloc{-(R-1)}{0}$.
Thus, the largest eigenvalue $\spsi$ is $1$, and the corresponding right eigenvector
has all its components equal.

On the opposite, the parametric form (\ref{psi}) where $\lambda_l$ are free parameters is in general
 \textit{not normalized}, with deep consequences discussed in the next sections. However,
there exists a transformation allowing to convert an arbitrary range-$R$ potential to a normalized potential
$\bPsi$ by the transformation:

\beq\label{normal}
\bPsi_{w'w}=\bpsi_{w'}-\log(\rpfwp)+\log(\rpfw) -\pres.
\eeq

Let us give two examples. First, if $\bpsi$ is normalized then $\rpfwp=\rpfw$
and $\pres=0$ so that (fortunately) $\bPsi=\bpsi$. Second,
if $\psi$ has range-$1$ then, according to the computations done in
section \ref{SlinkGibbs}, $\bPsi_{w'w}=-\log(Z)+\psi_{w'} = -\log(Z)+\psi(\bom(0))$.
Here, the normalisation only consists of removing the log of a (constant) partition function.

In the general case, the potential (\ref{normal}) has 
range-$R+1$. The corresponding RPF operator $L(\bPsi)$, is therefore the transition matrix 
for  a $R$-step Markov chain. Thus, switching to a parametric form (\ref{psi}) without
constraint on the $\lambda_l$'s we end up with a redundant transition
probability of form $P'(\omega(0) \, | \, \bloc{-R}{-1})$ while
the right transition probability is $P(\omega(0) \, | \, \bloc{-(R-1)}{-1})$.
Since, obviously $P'(\omega(0) \, | \, \bloc{-R}{-1})=P(\omega(0) \, | \, \bloc{-(R-1)}{-1})$
the final  form of the normalized potential can be easily simplified. 

\sssu{Probability of arbitrary spike blocks}\label{SPany}

Using the normalized potential (\ref{normal})
the probability
of a \textit{admissible} spike block of size strictly larger than $R$, $\bloc{t}{t+n+R-1}$, $t \in \setZ$, $n>0$ is given by:
$$\mpg\left[\bloc{t}{t+n+R-1}\right]
=\mpg\left[\bom(t),\bom(t+1) \dots \bom(t+n+R-1)\right]
=\mpg\left[w_t, \dots, w_{t+n}\right],$$
where the word $w_k$ encodes the block $\bloc{k}{k+R-1}$. As a consequence,
\nid 
\beq\label{ChapKol}
\mpg\left[w_t, \dots, w_{t+n}\right]= 
\mpg[w_{t}]
L_{w_{t},w_{t+1}}(\bPsi)
L_{w_{t+1},w_{t+2}}(\bPsi) \dots  
L_{w_{t+n-1},w_{t+n}}(\bPsi).
\eeq
 This is the classical
Chapman-Kolmogorov equation. Returning to the initial (non-normalised) potential
(\ref{psi}) this relation reads, using (\ref{normal}):
\beq\label{ChapKolNN}
\mpg\left[w_t, \dots, w_{t+n}\right]= 
\mpg[w_{t}]\frac{\rpfc{w_{t+n}}}{\rpfc{w_{t}}}
\frac{1}{e^{(n+1) \pres}}e^{\sum_{k=t}^{t+n} \psi_{w_{k}}}.
\eeq

One checks\footnote{Taking into account the fact that symbols $w_k$ encode spike
blocks of length $R$.} that $\mpg\left[w_t, \dots, w_{t+n}\right]$
satisfies the definition of a Gibbs distribution (\ref{defGibbs})
with $\pres=\log \spsi$ and $w_k=\sigma^k w_0$.

On the opposite, for blocks of size $0<n<R+1$ then 
$$\mpg\left[\bloc{t}{t+n}\right]= 
\sum_{w \subset \bloc{t}{t+n}} \mu_{\bpsi}(w),
$$
where the sum holds on each word $w$ containing the block $\bloc{t}{t+n}$.

\sssu{Links with the simple Gibbs form.}  \label{SlinkGibbs}
In this section we make the link between our formalism and previous approaches
using the simple Gibbs formulation.

As a preliminary remark note that the Gibbs-probability of a spike block $w$ of length $R$,
given by (\ref{invmeascomp}),
\textit{hasn't} the form $\frac{1}{Z}e^{\bpsi(w)}$, with $Z$ constant, \textit{except when $R=1$}.
The  case $R=1$ corresponds to a Markov
chain without memory, where therefore the spiking pattern $w_t=\bom(t)$ is independent on $w_{t-1}=\bom(t-1)$.
Examples are the Bernoulli model (where moreover spikes are spatially independent) or the
Ising model (where spikes are spatially correlated but not time correlated). In this case,
all transitions are allowed, thus the RPF matrix reads $\RPFww=e^{\bpsi_{w'}}$
and does not depend on $w$.
 As a consequence, all lines are linearly dependent which implies
that there are $N-1$ $0$-eigenvalues while the largest 
eigenvalue is 
$Z \deq \spsi=Tr(\RPF)=\D{\sum_{w' \in \cH} e^{\bpsi_{w'}}}$. 
The corresponding left eigenvector is $\lpf=(1, \dots, 1)$ and the right eigenvector is 
$\rpfc{w'}= \frac{e^{\bpsi_{w'}}}{Z}$, so that $\lpf.\rpf=1$.
 Thus, the Gibbs distribution is, according to (\ref{invmeascomp}),
$\mu_{w'}= \frac{e^{\bpsi_{w'}}}{Z}$. 

Consider now larger ranges. 
Recall first that a potential of form (\ref{psi}) is in general not normalized.
To associate it to a Markov chain one has to use the transformation (\ref{normal})
and the probability of a spiking pattern sequence is given by (\ref{ChapKolNN}).  
In particular, the joint probability of two admissible successive blocks $w',w$ 
reads $\mpg(w,w')=
\mpg[w']\frac{\rpfc{w}}{\rpfc{w'}}
\frac{1}{e^{\pres}}e^{\psi_{w'}}.
$
One can introduce a formal Hamiltonian $H_{ww'}=\bpsi_{w'}+\log(\rpfc{w})$ and
a ``conditional'' partition function $Z(w')=e^{\pres}\rpfc{w'}$ such
that $\mpg(w|w')=\frac{1}{Z(w')}e^{H_{ww'}}$ with $Z(w')=\sum_{w \in \cH} e^{H_{ww'}}$
but here the partition function depends on $w'$ (compare with eq. (1) in ref \cite{marre-boustani-etal:09}). 
This corresponds, in statistical mechanics, to have interactions with a boundary.
In this setting, the free energy density (topological pressure) is obtained (away from phase transitions\footnote{This
requires a sufficiently fast decay of the potential, as mentioned in the footnote \ref{Footpot}}), via   
$\frac{1}{n}\log Z_n(w') \to \pres$ as $n \to \infty$, $\forall w'$, requiring to consider a thermodynamic limit,
as we do in the present setting.

As a conclusion, starting from an a priori form of a potential (\ref{psi}), obtained e.g. by Jaynes
argument (see section \ref{SJaynes}) one obtains a non normalized potential which cannot be directly
associated with a Markov chain, and the corresponding Gibbs measure hasn't the simple Gibbs form
used for Ising model, as soon as one introduces memory terms in the potential. However, 
the thermodynamic formalism allows one to treat this case without approximations, or 
assumptions such as detailed balance, and
gives direct access to the topological pressure. 

\subsubsection{Comparing several Gibbs statistical models.}
The choice of a potential (\ref{psi}), i.e. the choice
of a set of observables, fixes a statistical model
for the statistics of spike trains. Clearly, there are many choices
of potentials and one needs to propose a criterion to compare them.   
The Kullback-Leibler divergence, 

\beq\label{HKL}
d(\nu,\mu)=\limsup_{n \to \infty} \frac{1}{n}\sum_{\Con} 
\nu\left(\Con\right)
\log\left[\frac{\nu\left(\Con\right)}{\mu\left(\Con\right)} \right],
\eeq
where $\nu$ and $\mu$ are two invariant probability measures,
provides some notion of asymmetric ``distance'' between $\mu$ and $\nu$.

The computation
of $d(\nu,\mu)$ is delicate but, in the present context,
the following holds.
For $\nu$ an invariant measure and $\mpg$ a Gibbs measure with a potential
$\bpsi$, both defined on the same set of sequences  $\Sigma$, one has 
\cite{bowen:98,ruelle:69,keller:98,chazottes-keller:09}:  
\beq\label{dKLpres}
d\left(\nu,\mpg \right) = \pres - \nu(\bpsi) - h(\nu).
\eeq
This is the key of the algorithm that we have developed. 


\subsection{Computing the Gibbs distribution from empirical data.}\label{Semp}

\subsubsection{Empirical Averaging}
Assume now that we observe the spike trains generated by the neural network.
We want to extract from these observations informations about the set of monomials $\phi_l$ constituting
the potential and the corresponding coefficients $\lambda_l$. 

Typically, one observes, from $\cN$ repetitions of the same experiment,
i.e. submitting the system to the same conditions,
 $\cN$ raster plots $\omega^{(m)}, m=1 \dots \cN$ on a finite time horizon 
of length $T$. These are the basic data from which we want
to extrapolate the Gibbs distribution. The key object for this
is the \textit{empirical} measure. For a fixed $\cN$ (number of observations)
and a fixed $T$ (time length of the observed spike train), the \textit{empirical
average} of a function $f: \Sigma \to \setR$ is:
\beq\label{empav}
\bar{f}^{(\cN,T)}=\frac{1}{\cN T}\sum_{m=1}^{\cN} \sum_{t=1}^T f(\sit \omega^{(m)}).
\eeq

 Typical examples  are $f(\omega)=\omega_i(0)$ in which case 
the empirical average of $f$ is the firing rate\footnote{Recall that we assume dynamics is stationary
so rates do not depend on time.} of neuron $i$;  
$f(\omega)=\omega_i(0)\omega_j(0)$ then the empirical average of $f$ measures the estimated 
probability of spike coincidence for neuron $j$ and $i$; $f(\omega)=\omega_i(\tau)\omega_j(0)$ then the empirical average of $f$ measures the estimated probability of 
the event ``neuron $j$ fires and neuron $i$ fires $\tau$ time step later'' (or sooner according to the sign
of $\tau$). 

Note that in (\ref{empav}) we have used the shift $\sit$ for the time evolution of the raster plot. This notation is
compact and well adapted to the next developments than the
classical formula, reading, e.g., for firing rates 
$\frac{1}{\cN T}\sum_{m=1}^{\cN} \sum_{t=1}^T f(\omega^{(m)}(t))$.

The empirical measure is the probability distribution $\pTo$ such that, for any function\footnote{In fact,
it is sufficient here to consider monomials.}
$f: \Sigma \to \setR$,
\beq\label{pTo}
\pTo(f)=\bar{f}^{(\cN,T)}.
\eeq

Equivalently, the empirical probability of a spike block $\bloc{t_1}{t_2}$ is given by:
\beq\label{empProb}
\pTo\left(\bloc{t_1}{t_2}\right)=\frac{1}{\cN T}\sum_{m=1}^{\cN} \sum_{t=1}^T \chi_{\bloc{t_1}{t_2}}(\sit \omega^{(m)}),
\eeq
where $\chi_{\bloc{t_1}{t_2}}$ is the indicatrix function of the block $\bloc{t_1}{t_2}$
so that $\sum_{t=1}^T \chi_{\bloc{t_1}{t_2}}(\sit \omega^{(m)})$ simply counts
the number of occurences of the block $\bloc{t_1}{t_2}$ in the empirical raster
$\omega^{(m)}$.

\sssu{Estimating the potential from empirical average}

The empirical measure is what we get from experiments
while it is assumed that spike statistics is governed
by an hidden Gibbs distribution $\mpg$ that we want to determine
or approximate. Clearly there are infinitely many \textit{a priori} choices
for this distribution, corresponding to infinitely many a priori choices for the potential
$\bpsi$. However, the ergodic theorem (the law of large number) states that
$\pTo \to \mpg$ as $T \to \infty$ where $\mpg$ is the sought Gibbs distribution.
Equivalently,
the Kullback-Leibler divergence $d\left(\pTo,\mpg \right)$ between the empirical measure
and the sought Gibbs distribution \textit{tends to} $0$ as $T \to \infty$.

Since we are dealing with finite samples the best that we can expect is to find
a Gibbs distribution which \textit{minimizes} this divergence. This is
the core of our approach. Indeed,  using eq. (\ref{dKLpres}) we use the approximation\footnote{This is an approximation because $\pTo$ is not invariant \cite{keller:98}. 
It becomes exact as $T \to +\infty$.}: 

\beq\label{KLemp}
d(\pTo,\mpg)=\pres-\pTo(\bpsi) -h(\pTo).
\eeq
The advantage is that this quantity can be numerically estimated,
since for a given choice of $\bpsi$ the topological pressure is known from the Ruelle-Perron-Frobenius theorem,
while $\pTo(\bpsi)$ is directly computable. Since $\pTo$ is fixed by the experimental raster plot,  
$h(\pTo)$ is independent of the Gibbs potential, so we can equivalently minimize: 
\beq \label{criterion_gen} 
\tilde{h}\left[\bpsi\right] = P\left[\bpsi\right]-\pTo(\bpsi),
\eeq
\textit{without computing the entropy $h(\pTo)$}.

This relation holds for any potential.
In the case of a parametric potential of the form (\ref{psi}) we have to minimize
\beq \label{criterion} 
\tilde{h}\left[\bl\right] = P\left[\bl\right]-\sum_{l=1}^L \lambda_l \pTo(\phi_l).
\eeq
Thus, from (\ref{Gener}) and (\ref{empav}), given the parametric form,
the set of $\lambda_l$'s minimizing the KL divergence are given by:
\beq\label{tangence}
\mpg\left[\phi_l\right]=\pTo(\phi_l), \qquad l=1 \dots L.
\eeq
Before showing why this necessary condition is also sufficient, we want to comment
this result in connection with standard approaches (``Jaynes argument'').

\subsubsection{Inferring statistics from empirical averages of observables: The Jaynes argument.} \label{SJaynes}

The conditions (\ref{tangence}) impose constraints on the sought Gibbs distribution.
In view of the variational principle (\ref{VarPrinc}) the minimization of KL
divergence \textit{for a prescribed parametric form of the Gibbs potential} is equivalent
to \textit{maximizing the statistical entropy under the constraints (\ref{tangence})},  where
the $\lambda_l$'s appear as adjustable Lagrange multipliers.
This is the Jaynes argument \cite{jaynes:57} commonly used to introduce Gibbs distributions
in statistical physics textbooks, and also used in the fundating paper of Schneidman et al. \cite{schneidman-berry-etal:06}.
There is however an important subblety that we want to outline. The Jaynes argument
provides the Gibbs distribution which minimizes the KL divergence with respect to the empirical
distribution \textit{in a specific class of Gibbs potentials}. Given a parametric
form for the potential it gives the set of $\lambda_l$'s which minimizes the KL divergence
for the set of Gibbs measures having \textit{this form of potential} \cite{csiszar:84}.  
Nevertheless, the divergence can still be quite large and the corresponding
parametric form can provide a poor approximation of the sought measure.
So, in principle one has to minimize the KL divergence with respect to
several parametric forms. This is a way to compare the statistical models.
The best one is the one which minimizes (\ref{criterion}), 
i.e. knowing if the `` model''$\bpsi_2$ is significantly ``better'' than $\bpsi_1$, reduces to verifying:
\begin{equation} \label{comparison} 
\tilde{h}\left[\bpsi_2\right] \ll \tilde{h}\left[\bpsi_1\right],
\end{equation}
easily computable at the implementation level, as developed below.
 Note that $\tilde{h}$
has the dimension of entropy. Since we compare entropies, which units are bits of information, defined in base 2,
the previous comparison units is well-defined.

\subsubsection{Convexity.} \label{convexity}

The topological pressure is convex with respect to $\bl$. 
As being the positive sum of two (non strictly) convex criteria $P\left[\bpsi\right]$ and $-\pTo(\bpsi)$ 
in ~(\ref{criterion}), the minimized criterion is convex. This means that the previous minimization method  intrinsically converges towards a global minimum.

Let us now consider the estimation of an hidden potential
$\psi=\sum_{l=1}^{L} \lambda_l \phi_l$ by a test potential 
$\psi^{(test)}=\sum_{l=1}^{L^{(test)}} \lambda^{(test)}_l \phi^{(test)}_l$.
As a consequence, we  estimate $\psi$ with a set of parameters
$\lambda^{(test)}_l$, and the criterion (\ref{criterion})
is minimized with respect to \textit{those parameters} $\lambda^{(test)}_l$, $l=1 \dots L^{(test)}$. 

Several situations are possible. 
First, $\psi$ and $\psi^{(test)}$ have the same
set of monomials, only the $\lambda_l$'s must be determined.
Then, the unique minimum
is reached for $\lambda^{(test)}_l=\lambda_l, l=1 \dots L$.
Second, $\psi^{(test)}$ contains all the monomials of $\psi$
plus additional ones (\textit{overestimation}).
 Then, the $\lambda_l^{(test)}$'s corresponding
to monomials in $\psi$ converge to $\lambda_l$ while the coefficients
corresponding to additional monomials converge to $0$.
The third case corresponds to \textit{underestimation}.
$\psi^{(test)}$ contains less monomials than $\psi$
or distinct monomials. In this case, there is still a minimum for the
criterion (\ref{criterion}), but it provides a statistical model
(a Gibbs distribution) at \textit{positive KL distance} from the correct potential \cite{csiszar:84}.
In this case adding monomials to $\psi^{(test)}$ will improve
the estimation. More precisely, if for a first 
test potential the coefficients obtained after minimisation of
$\tilde{h}$ are $\lambda^{(test)}_l, l=1 \dots L^{(test)}$
and for a second test potential they are $\lambda^{'(test)}_l, l=1 \dots L^{'(test)},  
L^{'(test)} >  L^{(test)}$ then $\tilde{h}(\lambda^{(test)}_1, \dots, \lambda^{(test)}_{L^{(test)}})
 \geq \tilde{h}(\lambda^{'(test)}_1, \dots, \lambda^{(test)}_{L^{'(test)}})$.
For the same $l$ the coefficients $\lambda^{(test)}_l$
and $\lambda^{'(test)}_l$ can be quite different.

Note that these different situations are not inherent to our procedure, but to the principle of finding an hidden probability by maximizing the statistical entropy under constraints, when the full set of constraints is not known\footnote{The problem of estimating the memory order of the underlying markov chain to a given sequence, which means, in our framework, to find the the potential range, has been a well known difficult question in coding and information theory \cite{morvai-weiss:08}. Some of the current available tests might offer additional algorithmic tools that would be explored in a forthcoming paper}.
Examples of these cases are provided in section \ref{Results}.
As a matter of fact, we have therefore two strategies to estimate
an hidden potential. Either starting from a minimal form of
test potential (e.g. Bernoulli) and adding successive monomials
(e.g. based on heuristical arguments such as ``pairwise correlations
do matter'') to reduce the value of $\tilde{h}$.
The advantage is to start from potentials with a few number coefficients,
but where the knowledge of the coefficients at a given step cannot
be used at the next step, and where one has no idea on ``how far''
we are from the right measure. The other strategy consists of starting
from the largest possible potential with range $R$ \footnote{\text{ibid.}}
. In this
case it is guarantee that the test potential is at the minimal
distance from the sought one, in the set of range-$R$ potential,
while the minimization will remove irrelevant monomials (their
coefficient vanishes in the estimation). The drawback is that one
has to start from a very huge number of monomials ($2^{NR}$) which reduces
the number of situations one can numerically handle. These two approaches
are used in section \ref{Results}.

\subsubsection{Finite sample effects and large deviations.}\label{SfiniteT}

Note that the estimations crucially depend on $T$. This is a central problem, 
not inherent to our approach but to all statistical methods where one tries to extract statistical properties from finite empirical sample. 
Since $T$ can be small in practical experiments, this problem can be circumvented by using an average over several samples (see eq. (\ref{empav}) and related comments).
Nevertheless it is important to have an estimation of finite sampling effects, which can be addressed
by the large deviations properties of Gibbs distributions.

 For each observable $\phi_l, \, l =1 \dots L$,
the following holds, as $T \to +\infty$ \cite{dembo-zeitouni:93}:
\beq \label{Conv_av_phil}
\mpg
\left\{ \omega \, , \, |\pTo(\phi_l)-\mpg\left(\phi_l \right)| \geq \epsilon \right\}
\sim \exp(-T I_l(\epsilon)),
\eeq
where $I_l(x)=\sup_{\lambda_l \in \setR} \left(\lambda_l x - P\left[\bl \right] \right)$,
is the Legendre transform of the pressure $P\left[\bl \right]$.

This  result provides the convergence rate with respect to $T$.
This is very important, since, once the Gibbs distribution is known,
one can infer the length $T$ of the time windows over which averages
must be performed in order to obtain reliable statistics. This is
of particular importance when applying statistical tests such as Neymann-Pearson
for which large deviations results are available in the case of Markov chains
and Gibbs distributions with finite range potentials \cite{negaev:02}.

Another important large deviation property also results from thermodynamic formalism \cite{keller:98,chazottes:99,dembo-zeitouni:93}. Assume that the experimental raster plot $\omega$
is distributed according to the Gibbs distribution $\mpg$, with  potential $\bpsi$, and assume that we propose, as a statistical model,
a Gibbs distribution with potential $\bpsi' \neq \bpsi$. 
 The Gibbs measure of spike blocks of range (\ref{invmeascomp}) is a vector
in $\cH$ and $\pTo=\left(\pTo(w)]\right)_{w \in \cH}$ is a random vector. Now, 
the probability $\mpg\left\{\|\pTo-\mu_{\bpsi'}\|<\epsilon\right\}$ that $\pTo$ is $\epsilon$-close
to the ``wrong'' probability $\mu_{\bpsi'}$ decays exponentially fast as,

\beq\label{LargeDev}
\mpg\left\{\|\pTo-\mu_{\bpsi'}\|<\epsilon\right\} \sim 
\exp(-T \inf_{\mu, \, \|\mu-\mu_{\bpsi'}\|<\epsilon} 
d(\mu,\mu_{\bpsi}))).
\eeq

Thus, this probability decreases exponentially fast with $T$,
with a rate given (for small $\epsilon$) by 
$T \, d(\mpg,\mu_{\bpsi'})$. Therefore, a difference of $\eta$ in the Kullback-Leibler
divergences  $d(\pTo,\mu_{\bpsi})$, $d(\pTo,\mu_{\bpsi'})$ leads to a ratio
$\frac{\mpg\left\{\|\pTo-\mu_{\bpsi}\|<\epsilon\right\}}
{\mpg\left\{\|\pTo-\mu_{\bpsi'}\|<\epsilon\right\}}$ of order $\exp{-T\eta}$. Consequently, for
$T \sim 10^8$ a divergence of order $\eta=10^{-7}$ leads to a 
ratio of order $\exp(-10)$. Illustrations of this are given in section \ref{Results}.

\subsubsection{Other statistics related to Gibbs distributions.}
The K-L divergence minimization can be completed with other standard criteria
for which some analytical results are available in the realm of Gibbs distributions
and thermodynamic formalism.
Fluctuations of monomial averages about their mean are Gaussian, since Gibbs distribution
obey a central limit theorem with a variance controlled by the second derivative
of $P(\bl)$. Then, using a $\chi^2$ test seems natural. Examples are given in section \ref{Results}.
In order to compare the goodness-of-fit (GOF) for probability distributions of spike blocks, we propose at the descriptive level the box plots tests. On the other hand, quantitative methods to establish GOF are numerous and can be classified in families of 'test Statistics', the most important being the Power-Divergence methods (eg. Pearson-$\chi^2$ test), the Generalized Kolmogorov-Smirnov (KS) tests (eg. the KS and the Watson-Darling test) and the Phi-Divergence methods (eg. Cramer-von Mises test)\cite{cressie-read:84,chiu-liu:09}. Finally, to discriminate 2 Gibbs measures one can use the Neyman-Pearson criteria
since large deviations results for the Neyman-Pearson risk are available in this case
\cite{negaev:02}.
In the present paper we have limited our analysis to the most standard
tests (diagonal representations, box plots, $\chi^2$).


%
%

\section{Application: parametric statistic estimation.}\label{Methods}


Let us now discuss how the previous piece of theory allows us to estimate, 
at a very general level,  parametric statistics of spike trains.

We  observe 
$N$ neurons during a stationary period of observation $T$, 
assuming that statistics is characterized by an unknown Gibbs potential of range $R$.
The algorithmic\footnote{The code is available at \nolinkurl{http://enas.gforge.inria.fr/classGibbsPotential.html} } procedure proposed here decomposes in three steps:

\ben
\item {\em Choosing a statistical model}, i.e. fixing the potential (\ref{psi})
(equivalently, the relevant monomials or ``observables'').

\item {\em Computing the empirical average of observables}, i.e. determine them from the raster, using eq. (\ref{empav}).

\item {\em Performing the parametric estimation}, i.e. use a variational approach to determine the Gibbs potential.
\een

Let us describe and discuss these three steps, and then discuss the design choices.

\subsection{Choosing a model: rate, coincidence, spiking pattern and more.}

\subsubsection{The meaning of  monomials.}

In order to understand the power of representation of the proposed formalism, 
let us start reviewing a few elements discussed at a more theoretical level in the previous section.

We start with a potential limited to a unique monomial.

\begin{itemize}

 \item If $\bpsi=\omega_i(0)$, 
its related average value measures the firing probability or {\em firing rate} of neuron $i$;

 \item If $\bpsi(\tom)=\omega_i(0) \, \omega_j(0)$, we now measure the probability of spikes coincidence for neuron $j$ and $i$,  
as pointed out at the biological level by, e.g, \cite{grammont-riehle:99} and developed by \cite{schneidman-etal:06};
 
 \item If $\bpsi(\tom)=\omega_i(\tau) \, \omega_j(0)$, we measure the probability of the event ``neuron $j$ fires and neuron $i$ fires $\tau$ time step later'' (or sooner according to the sign of $\tau$); in this case the average value provides\footnote{Substracting the firing rates of $i$ and $j$.}
 the {\em cross-correlation} for a delay $\tau$ and the auto-correlation for $i = j$;

  \item A step further, if, say, $\bpsi(\tom)=\omega_i(0) \, \omega_j(0) \, \omega_j(1)$, we now take into account triplets
of spikes in a specific pattern (i.e. one spike from neuron $i$ coinciding with two successive spikes from neuron $j$);

\end{itemize}

These examples illustrate the notion of ``design choice'': the first step of the method being to choose the ``question to ask'', i.e. what is to be observed over the data.
In this framework, this translates in: ``choosing the form of the potential''. Let us enumerate a few important examples.

\subsubsection{Taking only rate or synchronization into account: Bernoulli and Ising potentials.} 

Rate potential are range-$1$ potentials, as defined before.
Such models are not very interesting as such, but have two applications: they are
 used to calibrate and study some numerical properties of the present methods,
and they are also used  to compare the obtained conditional entropy with more sophisticated models.

Ising potentials have been introduced by Schneidman and collaborators in \cite{schneidman-berry-etal:06}, taking rate and synchronization
of neurons pairs, as studied in, e.g. \cite{grammont-riehle:99}. This form is justified
by the authors using the  Jaynes argument.

Let us now consider potentials not yet studied, up to our best knowledge, in the present literature.

\subsubsection{Taking rate and correlations into account: RPTD-$k$ potentials.}
This is a key example for the present study. 
On one hand, the present algorithmic was  developed to take not only Bernoulli or Ising-like potential into account, but a large class of statistical model, 
including a {\em general second order model} (redundant monomial being eliminated), 
i.e. taking rate, {\em auto-correlation} (parametrized by $\lambda_{i\tau}$) 
and {\em cross-correlation} (parametrized by $\lambda_{ij\tau}$) into account.

Being able to consider such type of model is an important challenge, because it provides a tool to analyze not only synchronization between neurons, 
but more general temporal relations (see e.g. \cite{diesmann-gewaltig-etal:99,grammont-riehle:99,bressloff-coombes:03} for important applications).

Let us now turn to a specific example related to the neuronal network dynamic analysis.

\subsubsection{Taking plasticity into account: ``STDP'' potentials}

In \cite{cessac-rostro-etal:09} we considered Integrate-and-Fire neural networks with Spike-Time Dependent Plasticity of type:

\beq\label{Rexample}
\Wij'=\epsilon
\left[\ld \Wij+  \frac{1}{T} \sum_{t=T_s}^{T+T_s} \omej(t) \sum_{u=-T_s}^{T_s} f(u) \, \omei(t+u)\right],
\eeq

\nid where $\Wij$ is the synaptic weight from neuron $j$ to neuron $i$,
$-1 <\ld <0$ a term corresponding to passive LTD, $T$ a large time, corresponding
to averaging spike activity for the synaptic weights update, and,

$$
f(x)=
\left\{
\baR{llll}
A_- e^{\frac{x}{\tau_-}}, \ &x <0, \quad A_- < 0;\\
A_+ e^{-\frac{x}{\tau_+}},\ &x >0, \quad A_+ > 0;\\
0, \ &x=0;
\eaR
\right.
$$

\nid with $A_-<0$ and $A_+>0$, is the STDP function as derived by Bi and Poo \cite{bi-poo:01}.
The shape of $f$  has been obtained from statistical extrapolations of experimental data.
$T_s \deq 2 \max(\tau_+,\tau_-)$ is a characteristic time scale.
We argued that this synaptic weight adaptation rule produces, 
when it has converged, spike trains distributed according to a Gibbs distribution with potential:

\beq\label{STDP}
\bpsi(\omega)=
\sum_{i=0}^N \lambda^{(1)}_i \omega_i(0)
+
\sum_{i=0}^{N-1}\sum_{j=0}^{N-1} \lambda^{(2)}_{ij}
 \sum_{u=-T_s}^{T_s}  f(u) \omei(0)\omej(u).
\eeq 

When considering a large number of neurons, it becomes difficult to compute and check numerically
this joint probability over the whole population. Here, we propose to
consider a subset $\cP_s$ of $N_s< N$ neurons. 
In this case, the effects of the rest of the population can be written
 as a bulk term modulating the individual firing rates and correlations of the observed population, leading
to a marginal potential of the form:

\beq\label{STDP2}
\bpsi_{\cP_s}(\omega)=
\sum_{i \in \cP_s} \lambda^{(1)'}_i \omega_i(0)
+
\sum_{i,j \in \cP_s0}\sum_{j=0}^{N-1} \lambda^{(2)}_{ij}
\sum_{u=-T_s}^{T_s}  f(u) \omei(0)\omej(u).
\eeq 

Here, the potential is a function of both past and future. A simple way to embed this potential in our framework, 
is to shift the time by an amount of $T_s$, using the stationarity assumption.

\subsubsection{The general case: Typical number of observed neurons and statistics range.}

The previous piece of theory allows us to take any statistics of range $R$, among any set of $N$ neurons into account. 
At the numerical level, the situation is not that simple, since it appears, as detailed in the two next sections, that 
both the memory storage and computation load are in $O(2^{NR})$, except if the grammar is very restrictive, and the possible spike pattern blocks very sparse. 
Hopefully, we are going to see that estimation algorithms are rather efficient, thus do not lead to a complexity larger than $O(2^{NR})$.

It is clear that the present limitation is {\em intrinsic} to the problem, since we have {\em at least}, for a statistics of range $R$, to count the number
of occurrences of blocks of $N$ neurons of size $R$, and there are (at most) $2^{NR}$ of them. 
Fastest implementations must be based on the {\em partial} observation of only a subset of, e.g., the most preeminent occurrences.

Quantitatively, we consider ``small'' values of $N$ and $R$, typically a number of neurons equal to $N \in \{1, \simeq 8\}$, 
and Markov chain of range $R = \{1, \simeq 16\}$, 
in order to manipulate quantities of dimension $N \leq 8$, and $R \leq 16$, and such that $N(R + 1) \leq 18$.

\subsection{Computing the empirical measure: prefix-tree construction.}\label{prefix-tree}

For one sample ($\cN=1$),the empirical probability (\ref{pTo}) of the block $\bloc{-D}{t}, \, -D < t \leq 0$ is given
by 
$$\pTo(\bloc{-D}{t})=\frac{\#\bloc{-D}{t}}{T}.$$
thus obtained counting the number of occurrence $\#\bloc{-D}{t}, -D < t \leq 0$  of the block $\bloc{-D}{t}$  in the sequence $\bloc{-T}{0}$.
Since we assume that dynamics is stationary
we have, 
$\pTo(\bloc{-D}{t}) = \pTo(\bloc{0}{t+D})$. 

We observe that the data structure size has to be of order $O(2^{NR})$ (lower if the distribution is sparse), but does not depends on $T$.
Since many distributions are sparse (not all blocks occur, because the distribution is constrained by a grammar), 
it is important to use a sparse data structure, without storing explicitly blocks of occurence zero.

Furthermore, we have to study the distribution at several ranges $R$ and it is important to be able to factorize these operations.
This means counting in one pass, and in a unique data structure, block occurrences of different ranges.

The chosen data structure is a tree of depth $R+1$ and degree $2^N$. The nodes at depth $D$ count the number of occurrences of each block $\bloc{-D+t}{t}$,
of length up to $D \leq R+1$\footnote{The code is available at \nolinkurl{http://enas.gforge.inria.fr/classSuffixTree.html}.}.
It is known (see, e.g., \cite{grassberger:89} for a formal introduction) that this is a suitable data structure 
(faster to construct and to scan than hash-tables, for instance) in this context. 
It allows to maintain a computation time of order $O(TR)$, which does not depends on the structure size.

\subsubsection{The prefix-tree algorithm.}

Since we use such structure in a rather non-standard way compared to other authors, e.g. \cite{grassberger:89,gao-kontoyiannis-etal:08}, we detail the method here.


We consider a spike train $\tom_{-T}^0$, where time is negative.
The prefix-tree data structure for the present estimation procedure is constructed iteratively. 

\ben
\item Each spiking pattern at time $t$, $\omega(t)$,  is encoded by an  integer $w(t)$.

\item This given, before any symbol has been received, we start with the empty tree consisting only of the root. 

\item Then suppose for $-D < t \leq 0$ that the tree $\cT(\bloc{-T}{t-1})$ represents $\bloc{-T}{t-1}$. 
One obtains the tree $\cT(\bloc{-T}{t})$ as follows:
\ben
\item  One starts from the root and takes
	 branches corresponding to the observed symbols $\omega(t-D+1)$, $\cdots$, $\omega(t)$.
\item If ones reaches a leaf before termination, one replaces this leaf by an internal node and extends on the tree. 
\item Each node or leaf has a counter incremented at each access, thus counting the number of occurrence 
$\#\bloc{-D}{t}, -D < t \leq 0$  of the block $\bloc{-D}{t}$  in the sequence $\bloc{-T}{0}$.
\een
\een

The present data structure  not only allows us to perform the empirical measure estimation over a period of time $T$, 
but can also obviously be used to aggregate several experimental periods of observation. 
It is sufficient to add all observations to the same data structure.

\subsubsection{Generalization to a sliding window.}

Though we restrict ourselves to stationary statistics in the present work, it is clear that the present mechanism can be easily generalized 
to the analysis of non-stationary data set, using a sliding window
considering the empirical measure in $[t, t + T[$, then $[t + 1, t + 1 + T[$, etc..
This is implemented in the present data structure by simply counting the block occurrences observed at time $t$ and adding the block occurrences observed at time $T$,
yielding a minimal computation load. The available implementation has already this functionality (see section \ref{SNonStat} for an example).

\subsection{Performing the parametric estimation}

In a nutshell, the parametric estimation reduces to minimizing~(\ref{KLemp}), by calculating the topological pressure 
$\pres \equiv P(\bl)$ using~(\ref{ConvVect}) and the related theorem. 
The process decomposes into the following steps.

\subsubsection{Potential eigen-elements calculation.}

It has been shown in the theoretical section that the Ruelle-Perron-Frobenius operator eigen-elements allows one to derive 
all characteristics of the probability distribution. 
Let us now describe at the algorithmic level how to perform these derivations.

\ben
 \item The first step is to calculate the right-eigenvector $\rpf$ of the $\RPF$ operator, associated to the highest eigenvalue, using a standard power-method series:

$$\begin{array}{rcl} s^{(n)} &=& \|\RPF \, \bv^{(n-1)}\| \\ \bv^{(n)} &=& \frac{1}{s^{(n)}} \,  \RPF \, \bv^{(n-1)} \\ \end{array}$$
where $\bv^{(n)}$ is the $n$-th iterate of an initial vector $\bv^{(0)}$ and $s^{(n)}$ is the $n$-th iterate of an initial real value $s^{(0)}$.
With this method the pair $(s^{(n)},\bv^{(n)})$ converges to $(\spsi, \rpf)$ as soon as $\bv^{(0)}$ is not orthogonal to $\rpf$. In our case, after some numerical tests, it appeared a good choice to 
either set $\bv^{(0)}$ to  an uniform value, or to use the previous estimated value of $\rpf$, if available. This last choice is going to speed up the subsequent steps of the 
estimation algorithm.

The key point, in this iterative calculation, 
is that $\RPF$ is (hopefully) a sparse $2^{NR} \times 2^{NR}$ matrix, as outlined in the section
\ref{SRPF}.
As a consequence calculating $\RPF \, \bv$ is a $O(2^{N+NR}) \ll O(2^{2\,NR})$ operation, making explicit
the grammar in the implementation. 

The required precision on $(\spsi, \rpf)$ must be very high, for the subsequent steps to be valid, even if the eigenvector dimension is huge (it is equal to $2^{NR}$), 
therefore the iteration must be run down to the smallest reasonable precision level ($10^{-24}$ in the present implementation).

We have experimented that between $10$ to $200$ iterations are required for an initial uniform step in order to attain the required precision (for $NR \in {2..20}$), 
while less than $10$ iterations are sufficient when starting with a previously estimated value. 

From this 1st step we immediately calculate:

\ben

\item The topological pressure $P\left(\bpsi\right) = \log(\spsi)$.

\item The normalized potential $\bPsi_w$ (this normalized potential is also stored in a look-up table).
This gives us the transition matrix, which can be used to generate spike trains distributed
according the Gibbs distribution $\mpg$ and used as benchmarks in the section \ref{Results}.

\een

\item The second step is to calculate the left eigenvector $\lpf$, this calculation having exactly the same characteristics as for $\rpf$.

From this 2nd step one immediately calculates:

\ben
\item The Gibbs probability of a block $w$ given by (\ref{invmeascomp}), from which probabilities of any block
can be computed (section \ref{SPany}).

\item The theoretical value of the observables average $\mpg(\phi_l)$, as given in~(\ref{avobservable}).

\item The theoretical value of the distribution entropy $h\left[\mpg\right]$, as given in~(\ref{CalchKS}).

\een

After both steps, we obtain all useful quantities regarding the related Gibbs distribution: probability measure, observable value prediction, entropy. 
These algorithmic loops are direct applications of the previous piece of theory and show the profound interest of the proposed framework: 
given a Gibbs potential, all other elements can be derived directly.

\een

\subsubsection{Estimating the potential parameters.}

The final step of the estimation procedure is to find the parameters $\bl$ such that the Gibbs measure 
 fits at best with the empirical measure.
We have discussed why minimizing~(\ref{KLemp}) 
is the best choice in this context. Since $h(\pTo)$ is a constant with respect to ${\bf \lambda}$,
it is equivalent to minimize
$\tilde{h}\left[\bpsi_{\mathbf{\lambda}}\right]$ eq. (\ref{criterion}), where $\mpg(\phi_l)$ is given by~(\ref{avobservable}.
Equivalently, we are looking for a Gibbs distribution
$\mpg$ such that $\frac{\partial P\left[\bpsi_{\bf \lambda}\right]}{\partial \lambda_l}=\pTo(\phi_l)$
which expresses that $\pTo$ is tangent to $P$ at 
$\bpsi_{\bf \lambda}$ \cite{keller:98}.

\subsubsection{Matching theoretical and empirical observable values.}

As pointed out in the theoretical part, the goal of the estimation is indeed to find the parameters $\bl$ for which 
theoretical and empirical observable values match. The important point is that this is exactly what is performed by the proposed method: 
minimizing the criterion until a minimum is reached, i.e. until the gradient vanishes corresponding to a point where $\mpg(\phi_l) = \pTo(\phi_l)$,
thus where theoretical and empirical observable values are equal. 
Furthermore, this variational approach provides an effective method to numerically obtain the expected result.

At the implementation level, the quantities $\pTo(\phi_l)$ are the empirical averages of the observables, i.e. the observable averages computed on the prefix tree. 
They are computed once from the prefix tree. 
For a given $\bl$, $P(\lambda)$ is given by step 1.a of the previous calculation, while $\mpg(\phi_l)$ is given by the step 2.b.
It is thus now straightforward\footnote{Considering a simple gradient scheme, there is always a $\epsilon^k > 0$, small enough for the series $\bl_l^{k}$ and $\tilde{h}^k$, defined by: 
\\ \centerline{$\bl_l^{k+1} = \bl_l^{k} + \epsilon^k \, \frac{\partial \tilde{h}}{\partial \bl_l}(\bl_l^{k})$} \\
\\ \centerline{$0 \leq \tilde{h}^{k+1} < \tilde{h}^{k},$}
to converge, as a bounded decreasing series, since:
\\ \centerline{$\tilde{h}^{k+1} = \tilde{h}^{k} - \epsilon^k \, \left|\frac{\partial \tilde{h}}{\partial \bl_l}\right|^2 + O\left((\epsilon^k)^2\right)$.}}
to delegate the minimization of this criterion to any standard powerful non-linear minimization routine.

 We have implemented such a mechanism using the GSL\footnote{The GSL \nolinkurl{http://www.gnu.org/software/gsl}
multi-dimensional minimization algorithms taking the criteria derivatives into account used here is the Fletcher-Reeves conjugate gradient algorithm, while other methods such as the Polak-Ribiere conjugate gradient algorithm, 
and the Broyden-Fletcher-Goldfarb-Shannon quasi-Newton method appeared to be less efficient (in precision and computation times) on the benchmarks proposed
in the result section. Anyway, the available code \nolinkurl{http://enas.gforge.inria.fr/classIterativeSolver.html} allows us to consider these three alternatives,
thus allowing to tune the algorithm to different data sets.} implementation of non-linear minimization methods.
We have also made available the GSL implementation of the simplex algorithm of Nelder and Mead which does not require the explicit computation of a gradient like in eq. ~(\ref{criterion}). This alternative is usually less efficient than the previous methods, 
except in situations, discussed in the next section, where  we are at the limit of the numerical stability. In such a case the simplex method is still 
working, whereas other methods fail.

\subsubsection{Measuring the precision of the estimation.}

Once the quantity $\tilde{h}\left[\bpsi\right]=P\left[\bpsi\right]-\pTo(\bpsi)$ (eq. (\ref{criterion})) has been minimized
the Kullback-Leibler divergence $d(\pTo,\mpg) = \tilde{h}\left[\bpsi\right] - h(\pTo)$ 
determines a notion of ``distance'' between the empirical measure $\pTo$
and the statistical model $\mpg$. Though it is not necessary to compute $d(\pTo,\mpg)$
for the comparison of two statistical models $\mpg,\mu_{\bpsi'}$, the 
knowledge of $d(\pTo,\mpg)$, even approximate, is a precious indication of the method precision.
This however requires the computation of $h(\pTo)$.

Though the numerical estimation of $h(\pTo)$ is a far from obvious subject,
we have implemented the entropy estimation using definitions ~(\ref{hKS}) and~(\ref{hKS2}). In order to interpolate the limit~(\ref{hKS2}), we have adapted an interpolation method from~\cite{grassberger:89} and used 
the following interpolation formula.
Denote by $h(\pTo)^{(n)}$ the entropy estimated from a raster plot of length $T$, considering cylinders of size $n$.
We use the interpolation formula 
$h(\pTo)^{(n)} \simeq h^\infty + \frac{k}{n^c}$, where $h^\infty, k, c > 0$ are free parameters,
with $h(\pTo)^{(n)} \to  h^\infty$, as $n \to +\infty$.
The interpolation formula has been estimated in the least square sense, calculating $h(\pTo)^{(n)}$ on the prefix-tree.
The formula is linear with respect to $h^\infty$ and $k$, thus has a closed-form solution with respect to these two variables.
Since the formula is non-linear with respect to $c$, an iterative estimation mechanism is implemented.

\subsection{Design choices: genesis of the algorithm.}

Let us now discuss in details the design choices behind the proposed algorithm. 

The fact that we have an implementation able to efficiently deal with higher-order dynamics is the result of computational choices and validations, important to report here,
in order for  subsequent contributor to have the benefit of this part of the work.

\subsubsection{Main properties of the algorithm.}\label{SMainProp}

~\\ {\bf Convexity.} As indicated in the section   \ref{convexity}
 there is a unique minimum of the criterion. However, if
 $ \psi^{(test)}$ contains monomials which are not in $\bpsi$, 
the procedure  converges but there is an indeterminacy in the $\lambda_l$'s corresponding to
exogenous monomials. The solution is not unique, there is a subspace of equivalent solutions.
The rank of the topological pressure Hessian is an  indicator of such a degenerate case. Note that these different situations are not inherent to our procedure, 
but to the principle of finding an hidden probability by maximizing the statistical entropy under constraints, when the full set of constraints is not known \cite{csiszar:84}.

~\\ {\bf Finite sample effects.}
As indicated in the section \ref{SfiniteT}  the estimations crucially depend on $T$. This is a central problem, 
not inherent to our approach but to all statistical methods where one tries to extract statistical properties from finite empirical sample. 
Since $T$ can be small in practical experiments, this problem can be circumvented by using an average over several samples. 
In the present thermodynamic formalism it is possible to have an estimation of the size of fluctuations as a function of the potential, using the central limit theorem and the fact that the variance of fluctuations is given by the second
derivative of the topological pressure. This 
is a further statistical test where the empirical variance 
can be easily measured and compared to the theoretical predictions.

~\\ {\bf Numerical stability of the method.}
Two factors limitate the stability of the method, from a numerical point of view. 

The first factor is that the RPF operator  is a function of the {\em exponential} of the potential $\bpsi = \sum_l \lambda_l \, \phi_l$. As a consequence, positive or negative values of $\bpsi$
yield huge or vanishing value of $L_{\bpsi}$, and numerical instabilities easily occurs. 

%

However, though numerical instabilities are unavoidable, the good news is that they are easily detected, 
because we have introduced a rather large set of numerical tests in the code:

\ben
\item Negligible values (typically lower than $10^{-4}$) are set to zero, implicitly assuming that they correspond to hidden transition in the grammar.
\item Huge value (typically higher than $10^{4}$) generate a warning in the code.
\item Several coherent tests regarding the calculation of the RPF eigen-elements are implemented: we test that the highest eigenvalue is positive 
(as expected from the RPF theorem), and that the left and right RPF related eigenvectors yield  equal eigenvalues, as expected; 
we also detect that the power-method iterations converge in less than a maximal number of iteration
(typically $2^{10}$). We never found this spurious condition during our numerical tests.
When computing the normalized potential~(\ref{normal}), we verify that the right eigenvalue is $1$ up to some precision, 
and check that the normal potential is numerically normalized (i.e. that the sum of probabilities is indeed $1$, 
up to some ``epsilon''). 
\een
 In other words, we have been able to use all what the piece of theory developed in the previous section makes available, to verify that the numerical estimation is valid.

The second factor of numerical imprecision is the fact that some terms $\lambda_l \, \phi_l$ may be negligible with respect to others, so that the numerical estimation
of the smaller terms becomes unstable with respect to the imprecision of the higher ones. This has been extensively experimented, as reported in the next section.

~\\ {\bf Relation with entropy estimation.}
The construction of a prefix-tree is also the basis of efficient entropy estimation methods \cite{grassberger:89,schurmann-grassberger:96}. 
See \cite{gao-kontoyiannis-etal:08} for a  comparative about entropy estimation of one neuron spike train (binary time series).
Authors numerically observed that the context-tree weighting methods \cite{london-etal:02} is seen to provide the most accurate results.
This, because it partially avoids the fact that using small word-lengths fails 
to detect longer-range structure in the data, 
while with longer word-lengths the empirical distribution is severely under-sampled, leading to large biases.
This statement is weaken by the fact that the method from \cite{schurmann-grassberger:96} is not directly tested in \cite{gao-kontoyiannis-etal:08}, 
although a similar prefix-tree method has been investigated.
 
 However the previous results are restrained to relative entropy estimation of ``one neuron'' whereas the analysis of entropy of a {\em group of neurons} is targeted
if we want to better investigate the neural code. In this case \cite{schurmann-grassberger:96} 
is directly generalizable to non-binary (thus multi-neurons) spike trains,
whereas the context-tree methods seems intrinsically limited to binary spike-trains \cite{london-etal:02},
and the numerical efficiency of these methods is still to be studied at this level.

 Here,  we can propose an estimation for  the KS-entropy from eq. (\ref{CalchKS}).
Clearly, we compute here the entropy of a Gibbs statistical model $\mpg$ while methods above try to compute this entropy from
the raster plot. Thus, we do not solve this delicate problem, but instead, propose a method to benchmark these methods from raster plots obeying a Gibbs statistics, where the Gibbs distribution approaches at best the empirical measures
obtained from experiments. 

\subsubsection{Key aspects of the numerical implementation.}

~\\ {\bf Estimating the grammar from the empirical measure.}

The grammar defined in~(\ref{grammar}) is  implemented as a Boolean vector indexed by $w$ and estimated by observing, in a prefix-tree of depth at least $R+1$,
whose blocks $\bloc{-R-1}{0}$ occur at least once (allowed transition).
We make therefore here the (unavoidable) approximation that unobserved blocks correspond to forbidden words
(actually, our implementation allows to consider that a block is forbidden if it does not
appear more than a certain threshold value). There is however, unless a priori information about the distribution is available, no better choice. 
The present implementation allows us to take into account such a priori information, for instance related to global time constraints on the network dynamics, 
such as the refractory period. See \cite{cessac-rostro-etal:09} for an extended discussion.

~\\{\bf Potential values tabulation.}

Since the implementation is anyway costly in terms of memory size, we have choosen to pay this cost but obtaining the maximal benefit of it and we 
used as much as possible tabulation mechanisms (look-up tables) in order to minimize the calculation load. All tabulations are based on the following binary matrix:
$$\Q \in \{0, 1\}^{L \times 2^{NR}},$$
with $\Q_{l,w} = \phi_l(\bloc{-R}{0})$,
where $w$ is given by (\ref{omtow}).
$\Q$ is  the matrix of all monomial  values, 
 entirely defined by the choice of the parameter dimensions $N$, $R$ and $D$. 
It  corresponds to a ``look-up table'' of each monomial values
where $w$ encodes $\bloc{-R}{0}$. Thus the potential (\ref{psi}) writes
$\psi_w = (\Q \, \bl)_{w}$. We thus store the potential exponential values as a vector and get 
values using a look-up table mechanism, speeding-up all subsequent computations.

This allows to minimize the number of operations in the potential eigen-elements calculation.

\subsubsection{Appendix: About other estimation alternatives.}

 Though what is proposed here corresponds, up to our best knowledge, to the best we can do to estimate a Gibbs parametric distribution in the present context, 
this is obviously not the only way to do it, and we have rejected  a few other alternatives, which appeared less suitable. For the completeness of the presentation,
it is important to briefly discuss these issues.\\

~\\{\bf Avoiding RPF right eigen-element's calculation.} 
In the previous estimation, at each step, we have to calculate step 1 of the RPF eigen-element's derivation for the criterion value calculation
and step 2 of the RPF eigen-element's derivation for the criterion gradient calculation. 
These are a costly $O(2^{N+NR})$ operations. 

One idea is to avoid step 2 and compute the criterion gradient numerically. We have explored this track: we have calculated
$\frac{\partial \tilde{h}}{\partial \lambda_l} \simeq \frac{\tilde{h}(\lambda_l + \epsilon) - \tilde{h}(\lambda_l - \epsilon)}{2 \, \epsilon}$ 
for several order of magnitude, but always found a poorer convergence (more iteration and a biased result) compared to using the closed-form formula.
In fact, each iteration is not faster, since we have to calculate $\tilde{h}$ at two points thus, to apply step 1, at least two times. 
This variant is thus to be rejected.

Another idea is to use a minimization method which does not require the calculation of the gradient: we have experimented this alternative using the simplex minimization
method, instead of the conjugate gradient method, and have observed that both methods correctly converge towards a precise solution in most cases, while the 
conjugate gradient method is faster. However, there are some cases with large range potential, or at the limit of the numerical stability where the simplex method may still
converge, while the other does not.\\

~\\{\bf Using a simple Gibbs form.}
Using the Gibbs form 
$$ \mpg\left[w_t, \dots, w_{t+n}\right]= 
\frac{e^{\sum_{k=t}^{t+n} \psi_{w_{k}}}}{Z_n},
 \mbox{ with } Z_n = \sum_{w_t, \dots, w_{t+n}} e^{\sum_{k=t}^{t+n} \psi_{w_{k}}},$$
where $Z_n$ is  a constant,
could provide an approximation of the right Gibbs distribution and of the topological pressure,
 avoiding the power-method internal loop.
Furthermore, instead of a costly $O(2^{N+NR})$ operation, calculating $Z_n$ (and derivatives) 
 would require a simple scan of the prefix-tree (since values are calculated at each step
weighted by the empirical measure values) thus $O(2^{NR})$ operations. This apparent gain is 
unfortunately impaired since the amount of calculation is in fact rather heavy.
Moreover, as widely commented on section 2, the result is biased with a {\em non negligible}  additional bias  
increasing with the range $R$ of the potential. Finally, it has been observed as being slower than for the basic method.\\

~\\{\bf About analytical estimation of the RPF eigen-element's.}
The costly part of the RPF eigen-element's computation is the estimation of the highest eigenvalue. It is well-known that if the size of the potential is 
lower than five, there are closed-form solutions, because this problem corresponds to finding the root of the operator characteristic polynomial.
In fact, we are going to use 
this nice fact to cross-validate our method in the next section. However, except for toy's potentials (with $2^{NR} < 5 \Leftrightarrow NR \leq 2$ !), there is no chance 
that we can not do better than {\em numerically} calculating the highest eigenvalue. And the power method is known as the most powerful way to do it, in the general case.
We thus have likely optimal choices at this stage.\\

~\\{\bf Using other approximations of the KL-divergence criterion.}
Let us now discuss another class of variants: the proposed KL-divergence criterion in~(\ref{HKL}) and its empirical instantiation in~(\ref{KLemp}) are not the only one
numerical criterion that can be proposed in order to estimate the Gibbs distribution parameters. For instance, we have numerically explored approximation of the
KL-divergence of the form: 
$$d(\nu,\mu) \simeq \sum_{n = R}^{R'} \frac{\alpha_n}{n}\sum_{\Con} \nu\left(\Con\right) \log\left[\frac{\nu\left(\Con\right)}{\mu\left(\Con\right)} \right], $$
and have obtained coherent results (for $\alpha_n = 1$), but not quantitatively better than what is observed by the basic estimation method, 
at least for the set of performed numerical tests.

All these variants correspond to taking into account the same kind of criterion, but some other weighted evaluations of the empirical average of the observable. 
There is no reason to use it unless some specific a priori information on the empirical distribution is available. \\

Another interesting track is to use~(\ref{normal}) which allows us to write a KL-divergence criterion, 
not on the probability block, but on the conditional probability block,
as proposed in  \cite{chazottes:99,chazottes-etal:98} in a different context.
  We have considered this option. However a straightforward derivation allows one to verify, 
that this in fact corresponds the same class of criterion but with a different empirical observable average estimation. 
At the numerical level, we did not observe any noticeable improvement. \\

~\\{\bf Using score matching based estimation.}
We are here in a situation where we have to estimate a parametric statistical distribution, 
whose closed-form is given up to a scale factor $Z_n$. 
Such model contains a normalization constant whose computation may be considered as too difficult for practical purposes,
as it is the case for some maximum likelihood estimations. 
Score-matching methods \cite{hyvarinen:05} are based on the gradient of the log-density with respect to the data vector, 
in which the normalization constant is eliminated. 
However, the estimation criterion is no more the KL-divergence, and there is no guaranty that the obtained solution is not biased with 
respect to a well-defined statistical quantity.
As such it is another candidate to estimate Gibbs distribution.
However, thanks to the eigen decomposition of the RPF operator, we do not need to use this trick, since we obtain a tractable calculation of the normalization constant
at each step of the estimation and can minimize a well-defined criterion, as proposed in this paper. 

 We have numerically checked such modification of the criterion in which we do not consider the KL-divergence criterion, but the {\em ratio} between 
two conditional probabilities, as defined in~(\ref{normal}). Considering this ratio allows to eliminate the scale factor $Z_n$. 
This is the same spirit as score matching based estimation, more precisely, it corresponds to a discrete form of it, where the gradient of the log-density
is replaced by finite difference.
We have obtain correct results for simple forms of potential, 
but have experimented that the method is numerically less robust than using the unbiased method developed in this paper. 
This confirms that using the eigen-decomposition of the RPF operator, is the key for numerically stable estimations of such parametric statistics.\\
 
~\\{\bf Estimation in the case of a normalized potential.}
In the case where the potential is normalized, the criterion~(\ref{criterion}) is a simple linear criterion, thus unbounded and its minimization is meaningless.
In this singular case, its is obvious to propose another criterion for the estimation of the parameters.
A simple choice is to simply propose that the theoretical likelihood of the measure matches the estimated one, in the {\em least square sense}. 
This has been integrated in the available code.

\section{Results}\label{Results}

\subsection{Basic tests: validating the method}

\subsubsection{Method}

Knowing the potential $\bpsi$, it is easy to generate a spike train of length $T$, 
distributed according to $\mpg$, using the Chapman-Kolmogorov equations
(\ref{ChapKol}). Thus, we have considered several examples of Gibbs potentials,
where, starting from a sample raster plot $\bloc{-T}{0}$
distributed according to $\mpg$, we use our algorithm to  recover the right form of $\bpsi$. 

Given a potential of range-$R$ of the parametric form (\ref{psi})
 and a number of neurons $N$ we apply the following method:
\ben
\item Randomly choosing the parameter's values $\lambda_l, \, l=1 \dots L$ of the Gibbs potential;
\item Generating a spike train realization of length $T$;
\item From these values  re-estimating a Gibbs potential:
\ben
\item Counting the block occurrences, thus the probabilities $\pTo$ from the  prefix-tree,
\item Minimizing~(\ref{criterion}), given $\pTo$,  as implemented by the proposed algorithm.
\item Evaluating the precision of the estimation as discussed in the previous section.
\een
\een

In the previous method, there is a way to simulate ``infinite'' ($T = +\infty$) sequences, by skipping step 2., 
and  filling the prefix-tree in step 3.a directly by the exact probability $\mpg(w)$ of the 
blocks $w$.\\

At a first glance, this loop seems to be a ``tautology'' since we re-estimate the Gibbs potential parameters from a one-to-one numerical process. However, this is not the case for three reasons:

\ben
\item For $T = +\infty$ using the same potential for the prefix-tree generation 
and for the parameters estimation, must yield the same result, 
but \textit{up to the computer numerical precision}. This has to be controlled due
 to the non-linear minimization loop in huge dimension. 
This is obviously also a way to check that the code has no mistake. 

\item For $T < +\infty$ using the same potential allows us to study the numerical precision of the estimation in the realistic situation of finite size data set, 
providing quantitative estimations about the truncation effects to be expected.

\item Using different potentials between simulated data generation and the parameters value estimation allows us to study numerically to which extends we can only 
correctly estimate the parameter's values, even if huge state vectors are involved. Quantitative errors are obtained. 
We can also perform comparison between different statistical models, as detailed in the sequel.


\een

\subsubsection{An illustrative example to understand what the algorithm calculates}


Let us start with very simple example, for which we can make explicit what the algorithm calculates, thus helping
 the reader to understand in details what the output is.

We consider a situation where 
 the number $L$ of parameters $\lambda_l$ is known (only the values
of the $\lambda_l$'s are unknown). We start from rather basic examples and then increase their complexity. In the first examples analytical expression for
the topological pressure, entropy, RPF eigen-vectors and invariant measure are available. Thus we can check that we re-obtain, 
from the estimation method, the related values up to the numerical imprecision.

~\\ {\bf One neuron and range-$2$.}\label{S_calc_an_1}
Here $\bpsi(\omega)=\lambda_1\,\omega_0(0)+ \lambda_2\,\omega_0(0)\,\omega_0(1)$.
We obtain analytically:
%
$$
\baR{lll}
\spsi&=&\frac{1+B+\sqrt{(1-B)^2+4A}}{2},\\
\pres&=&\log\spsi,\\
\lpf&=&\left(1,\spsi-1,A,B(\spsi-1),\right)\\
\rpf&=&\left(\spsi-B,\spsi-B,1,1\right)^T,\\
\mpg&=&\frac{1}{\spsi^2+A-B}\left(\spsi-B,A,A,B(\spsi-1)\right),\\
h\left[\mpg \right]&=&
\log(\spsi)-\lambda_1\frac{\partial \spsi}{\partial \lambda_1}-\lambda_2\frac{\partial \spsi}{\partial \lambda_2}\\
r&=&\frac{A+B(\spsi-1)}{\spsi^2+A-B},\\
C&=&\frac{B(\spsi-1)}{\spsi^2+A-B},\\
\eaR,
$$
%
with $A=e^{\lambda_1}=e^{\psi_{10}}, B=e^{\lambda_1+\lambda_2}=e^{\psi_{11}}$ and where
$T$ denotes the transpose. We remind that the index vector encodes spike blocs
 by eq. (\ref{omtow}). Thus, the first index ($0$)
corresponds to the bloc $00$, $1$ to $01$, $2$ to $10$ and $3$ to $11$.
$r$ is the firing rate, $C$ the probability that the neuron fires two successive time steps.
This is one among the few models for which a closed-form solution is available. 

The following numerical verifications have been conducted.
A simulated prefix-tree whose nodes and values has been generated  using~(\ref{psi}) 
with $\lambda_1=\log(2), \lambda_2= \log(2)/2$.
We have run the estimation program of $\lambda_i$'s and have obtained the right values with a precision better than $10^{-6}$. We also obtain a precision better than $10^{-6}$ for $\spsi, r, C, h\left[\mpg \right]$.
This first test simply states that the code has no mistake. 

\begin{figure}[h!]
\label{simple-model-comparison}
\begin{center} 
\includegraphics[height=6cm,width=8cm]{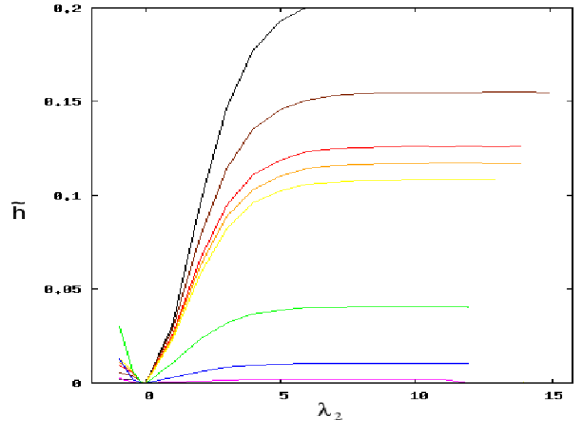} 
\end{center}
\caption{\footnotesize{ Entropy difference, using $\tilde{h}$, defined in~(\ref{criterion}), 
between the estimations of a range-$1$ and a range-$2$ model. The range-$2$ model writes
$\phi = -\lambda_1 \, \omega_0(0) - \lambda_2 \omega_0(0) \, \omega_0(1)$ for 
$\lambda_1 = \{ -1 \; (black), -0.5 \; (brown), -0.2 \; (red), -0.1 \; (orange), 0 \; (green), 1 \; (blue), 2 \; (Magenta)\}$.
$\lambda_2$ is a free parameter, in abscissa of this curve.
 The range-$1$ corresponds to $\lambda_2 = 0$.}}
\end{figure}

A step further, we have used this simple potential to investigate to which extends we can detect if the model is of range-$1$ (i.e. with $\lambda_2 = 0$) or range-$2$
(i.e. with a non-negligible value of $\lambda_2$). 
To this purpose, we have generated a range-$2$ potential and have performed its estimation using a range-$1$ and a range-$2$ potential,
 comparing the entropy difference 
(Fig.~\ref{simple-model-comparison}).


As expected the difference is zero for a range-$2$ model when $\lambda_2 = 0$, and this difference increases with 
 $\lambda_2$.
Less obvious is the fact that curves saturate for high values of $\lambda_2$. Indeed, 
because of the exponential function, 
high values of $\lambda$ yield huge or vanishing values of the RPF operator, thus numerical instabilities. 
This instability is detected by our algorithm.
Note that values of $\lambda$ larger than $10$
in absolute value have little sense from a 
statistical analysis of spike trains perspective.

We also have generated a range-$1$ potential and have performed its estimation, using a range-$1$ versus a range-$2$ model, and found  always that using range-$2$ model is as good as using a model of range-$1$ (not shown).

~\\ {\bf Two neurons and range-$2$ (Ising).}
Here $\bpsi(\omega)=\lambda_1\,\omega_1(0)+\lambda_2\,\omega_2(0)+ \lambda_3\,\omega_1(0)\,\omega_2(0)$.
The largest eigenvalue of the RPF operator is $Z=\spsi=A+B+C+D$,
with $A=1,B=e^{\lambda_1}, C=e^{\lambda_2},D=e^{\lambda_1+\lambda_2+\lambda_3}$ and the topological pressure
is $\log\spsi$. Here the Gibbs distribution has the classical form.
 We still obtain numerical precision better than $10^{-4}$, for 
standard values of $\lambda$, e.g., $\lambda_1=1, \lambda_2=\log(2), \lambda_3= \log(2)/2$.


~\\ {\bf Two neurons and pattern of spikes.}
A step further, we have considered $\bpsi(\omega)=\lambda_1\,\omega_1(0)+\lambda_2\,\omega_2(0)+ \lambda_3\,\omega_1(0)\,\omega_2(1)\,\omega_1(2)$, 
and $\bpsi(\omega)=\lambda_1\,\omega_1(0)+\lambda_2\,\omega_2(0)+ \lambda_3\,\omega_1(0)\,\omega_2(1)\,\omega_2(2)\,\omega_3(3)$, 
for random values drawn in $]-1, 0[$, i.e., considering the statistical identification of {\em spike patterns}.
We still obtain numerical precision better than $10^{-3}$, for these standard values of $\lambda$, though the precision decreases with the number of degrees of freedom,
as expected, while it increases with the observation time. This is investigated in details in the remainder of this section.\\

When considering larger neuron  $N$ and range-$R$ the main obstacle toward analytical results
is the Galois theorem which prevent a general method for the determination of the largest eigenvalue of the RPF operator. Therefore, we only provide numerical results obtained for more general potentials.

\subsubsection{Gibbs potential precision paradigm: several neurons and various ranges.} 

In order to evaluate the numerical precision of the method, we have run the previous benchmark considering potentials with all monomial of degree less or equal to 1,
and less or equal to 2, at a various ranges, with various numbers of neurons. 
Here we have chosen $T = +\infty$ and used the same potential for the prefix-tree generation and for the parameters value estimation. The computation time is reported
in Table~\ref{cpu-time} and the numerical precision in Table~\ref{divergence}, for $NR \leq 16$. This benchmark allows us to verify that there is no ``surprise''
at the implementation level: computation time increases in a supra-linear way with the potential size, but, thanks to the chosen estimation method, remains tractable in the size range compatible with available memory size. This 
 is the best we can expect, considering the intrinsic
numerical complexity of the method. Similarly, we observe that while the numerical precision decreases when considering large size potential, 
the method remains stable. Here tests has been conducted using the standard 64-bits arithmetic, while the present implementation can easily be recompiled using
higher numerical resolution (e.g. ``long double'') if required.

A step further, this benchmark has been used to explore the different variants of the estimation method discussed in the previous section
(avoiding some RPF eigen-element's calculation, using other approximations of the KL-divergence criterion, ..) and fix the details of the proposed
method.

\begin{table}[h]
\caption{\footnotesize{\label{cpu-time} Cpu-time order of magnitude in second ({\small using Pentium M 750 1.86 GHz, 512Mo of memory}), for the estimation of 
a potential with all monomial of degree less or equal to 1, for $\bpsi_1$ and less or equal to 2, for $\bpsi_2$, 
(i.e., $\bpsi_1(\tom)= \sum_{i=0}^{N-1} \lambda_i \omega_i(0)$, 
$\bpsi_2(\tom) = \sum_{i=0}^{N-1} \lambda_i \omega_i(0) + \sum_{i=0}^{N-1}\sum_{j=0}^{i-1} \sum_{\tau = -T_s}^{T_s} \lambda_{ij\tau}  \omega_i(0) \omega_j(\tau)$)
at a range-$R=2 T_s+1$ with $N$ neurons. 
We clearly observe the exponential increase of the computation time.
Note that the present implementation is not bounded by the computation time, 
but simply by the exponential increase of the memory size.}}
 {\small
\begin{tabular}{cc}
\begin{tabular}{l|ccccc}
\hline\noalign{\smallskip}
$\bpsi_1$  &R=1&R=2&R=4&R=8&R=16\\
\noalign{\smallskip}\hline\noalign{\smallskip}
 N=1& 2.0e-06 & 3.0e-06 & 8.0e-06 & 7.8e-05 & 2.9e-01 \\
 N=2& 4.0e-06 & 1.0e-06 & 3.0e-05 & 6.7e-02 & \\
 N=4& 1.3e-05 & 3.8e-05 & 8.3e-02 & & \\
 N=8& 2.4e-03 & 3.2e-01 & & & \\
\noalign{\smallskip}\hline
\end{tabular} 
\begin{tabular}{l|ccccc}
\hline\noalign{\smallskip}
  $\bpsi_2$  &R=1&R=2&R=4&R=8&R=16\\
\noalign{\smallskip}\hline\noalign{\smallskip}
 N=1& 4.5e-16 & 4.0e-06 & 4.0e-06 & 7.2e-04 & 3.7e-02 \\
 N=2& 3.0e-06 & 5.0e-06 & 4.0e-04 & 1.1e+00 & \\
 N=4& 1.9e-05 & 1.2e-03 & 3.6e+00 & & \\
 N=8& 6.6e-03 & 6.2e-01 & & & \\
\noalign{\smallskip}\hline
\end{tabular}
\end{tabular}
} 
\end{table}

\begin{table}[h]
\caption{\footnotesize{\label{divergence} Numerical precision of the method using synthetic data, for the estimation of $\bpsi_1$ and $\bpsi_2$, at a range-$R$ with $N$ neurons. 
The Euclidean distance $|\bar{\bf \lambda} - \tilde{\bf \lambda}|$ between the estimated parameter's value $\tilde{\bf \lambda}$ and the true parameter's value $\bar{\bf \lambda}$
is reported here, when the $\bar{\lambda}_l$'s are randomly drawn in $[-1,1]$.
We clearly observe the error increase, but the method remaining numerically stable.}}
{\small
\begin{tabular}{cc}
\begin{tabular}{l|ccccc}
\hline\noalign{\smallskip}
  $\bpsi_1$  &R=1&R=2&R=4&R=8&R=16\\ 
\noalign{\smallskip}\hline\noalign{\smallskip}
 N=1& 5.0e-09 & 2.2e-02 & 6.3e-03 & 1.3e-02 & 6.9e-03 \\
 N=2& 1.1e-08 & 1.3e-02 & 9.2e-03 & 5.2e-03 & \\
 N=4& 8.0e-09 & 8.5e-03 & 6.8e-03 & & \\
 N=8& 3.8e-08 & 5.1e-03 & & & \\
\hline\noalign{\smallskip}
\end{tabular} 
\begin{tabular}{l|ccccc}
\hline\noalign{\smallskip}
  $\bpsi_2$  &R=1&R=2&R=4&R=8&R=16\\ 
\noalign{\smallskip}\hline\noalign{\smallskip}
 N=1& 1.1e-10 & 1.9e-02 & 7.2e-03 & 4.8e-03 & 9.2e-02 \\
 N=2& 1.1e-09 & 4.8e-03 & 3.7e-03 & 2.3e-03 & \\
 N=4& 3.7e-08 & 2.6e-03 & 5.8e-02 & & \\
 N=8& 6.0e-06 & 2.4e-02 & & & \\
\hline\noalign{\smallskip}
\end{tabular}
\end{tabular} 
}
\end{table}

\subsection{More general tests: applying the method}

\subsubsection{Test framework.}

In order to test more general potentials for $N=2$ neurons we  explicit here the forms 
~(\ref{Ising}),~(\ref{RPTD-k}),~(\ref{PTD-k}), where $k \in \mathbb{N}$:
\begin{equation}\label{pot2}
\begin{split}
\text{Ising} : \bpsi(\omega) &=\lambda_1\,\omega_1(0)+\lambda_2\,\omega_2(0)+ \lambda_3\,\omega_1(0)\,\omega_2(0).\\
\text{RPTD}-k : \bpsi(\omega)&=\lambda_1\,\omega_1(0)+\lambda_2\,\omega_2(0)+\sum_{i=-k}^{i=k} \hat{\lambda}_i\,\omega_1(0)\omega_2(i).\\
\text{PTD}-k  : \bpsi(\omega)&=\sum_{i=-k}^{i=k} \hat{\lambda}_i\,\omega_1(0)\omega_2(i).\\
\end{split}
\end{equation}

\paragraph{test 1 (estimation precision).} Given a selected  potential of form ~(\ref{pot2}) we choose randomly its coefficients $\bar{\lambda}_l$ from an uniform distribution on $[-2,0]$ and we generate a spike-train of length $T=4 \times 10^8$.  Then we construct a prefix-tree from a sample of length $T_0 \ll T$ (typically $T_0=10^7$)
taken from the generated spike-train.
 For each  sample of length $T_0$ we propose a randomly chosen set of ``initial guess'' coefficients, used
to start the estimation method, distributed according to $\tilde{\lambda}^{(0)}_l=\bar{\lambda_l} (1+(U[0,1]-0.5)x/100)$, where $x$ is the initial percentage of bias from the original set of generating coefficients and
$U[0,1]$ is a uniform random variable on $[0,1]$.  Call $\tilde{\lambda}_l$ the values obtained after convergence of the algorithm. Results show that:

\ben
\item[(i)] the error $E(|\tilde{\lambda_l}-\bar{\lambda_l}|)$ increases with the range of the potential and it decreases with  $T_0$; 

\begin{figure}\label{test1f}
\begin{center}
 \includegraphics[height=7.5cm, width=8.5cm]{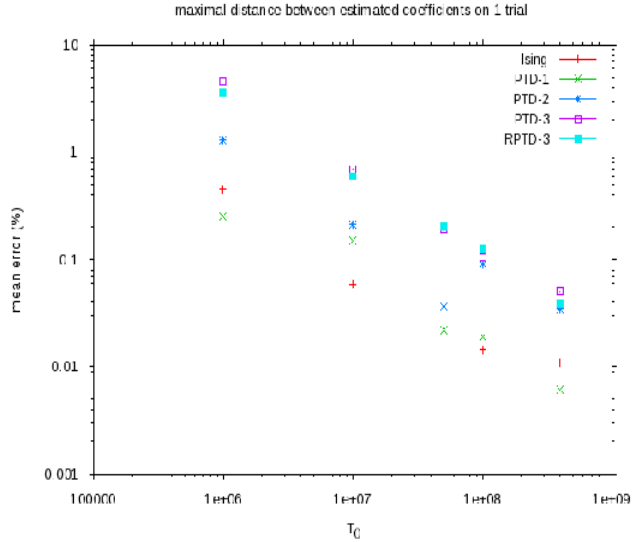}
\caption{\footnotesize{Mean error (in percentage) vs $T_0$ size.}}
\end{center}
\end{figure}

\item[(ii)] the error is independent of the initial bias percentage (see figs \ref{test1f}); 

\item[(iii)] $\tilde{h}\left[\bpsi\right]=P[\bpsi]-\pTo(\bpsi)$ is fairly constant with respect to
the length $T_0$ (not shown).
\een

\paragraph{Test 2 (Models comparison).}
We select a potential form $\bpsi$ from those proposed in ~(\ref{pot2}); we choose randomly its coefficients $\bar{\lambda}_l$ from an uniform distribution in $[-2,0]$; we generate a spike-train of length $T=1 \cdot 10^8$ and we construct the prefix-tree with the spike-train obtained. Using this prefix-tree we estimate the coefficients 
$\lambda^{\bpsi_m}_i$ that minimizes the KL divergence for several 
statistical models $\bpsi_m$ proposed in ~(\ref{pot2}).  The coefficients $\lambda^{\bpsi_m}_i$ and $\tilde{h}=P[\bpsi_m]-
\pTo\left(\bpsi_m\right)$ are averaged over $20$ samples and error bars are computed. Results show that :

\begin{enumerate}[(i)]
 
\item The 'best' statistical models (i.e the ones with lowest mean value KL divergence)  have the same monomials as the statistical model that generated the spike-train, plus, possibly additional monomials. For example, in ~(\ref{pot2}), \textbf{RPTD-$1$} contains \textbf{Ising}, and also the \textbf{PTD-$1$} but not 
\textbf{PTD-$2$}. We choose the model with  the minimal number of coefficients 
in agreement with section \ref{convexity}.

\item The value of the additional coefficients of an over-esimated model 
(corresponding to monomials absent in the corresponding potential) are almost null up to the numerical error.

\item For all the 'best' suited statistical models (in the sense of (i)), the criterion $\tilde{h}\left[\bpsi\right]$ ~(\ref{criterion}) averaged over trials, is fairly  equal for these models up to a difference of order $\delta \approx 10^{-6}$, and the difference with respect to other types of statistical models  is at least of $4$ orders of magnitude lower.
We recall that, according to section ~\ref{SfiniteT},
 the deviation probability is of order
to $ \exp(-\delta T)$. After estimation from a raster generated with an Ising model, the ratio of the deviation probabilities  ~(\ref{LargeDev})  between an  \textbf{Ising} and a \textbf{RPTD-$1$} model  is 
$\sim \eta=\exp(0.0000115 \times 10^8)$  , while between the \textbf{Ising} and the \textbf{PTD-$3$}  
$\sim \eta= \exp(0.00072 \times 10^8)$ meaning that the \textbf{PTD-$3$} provide a worst estimation.

\item  The predicted probability of words corresponds very well with the empirical value.
\end{enumerate}

\begin{figure}\label{test2x}
\begin{center}
 \includegraphics[height=8.0cm, width=8.5cm]{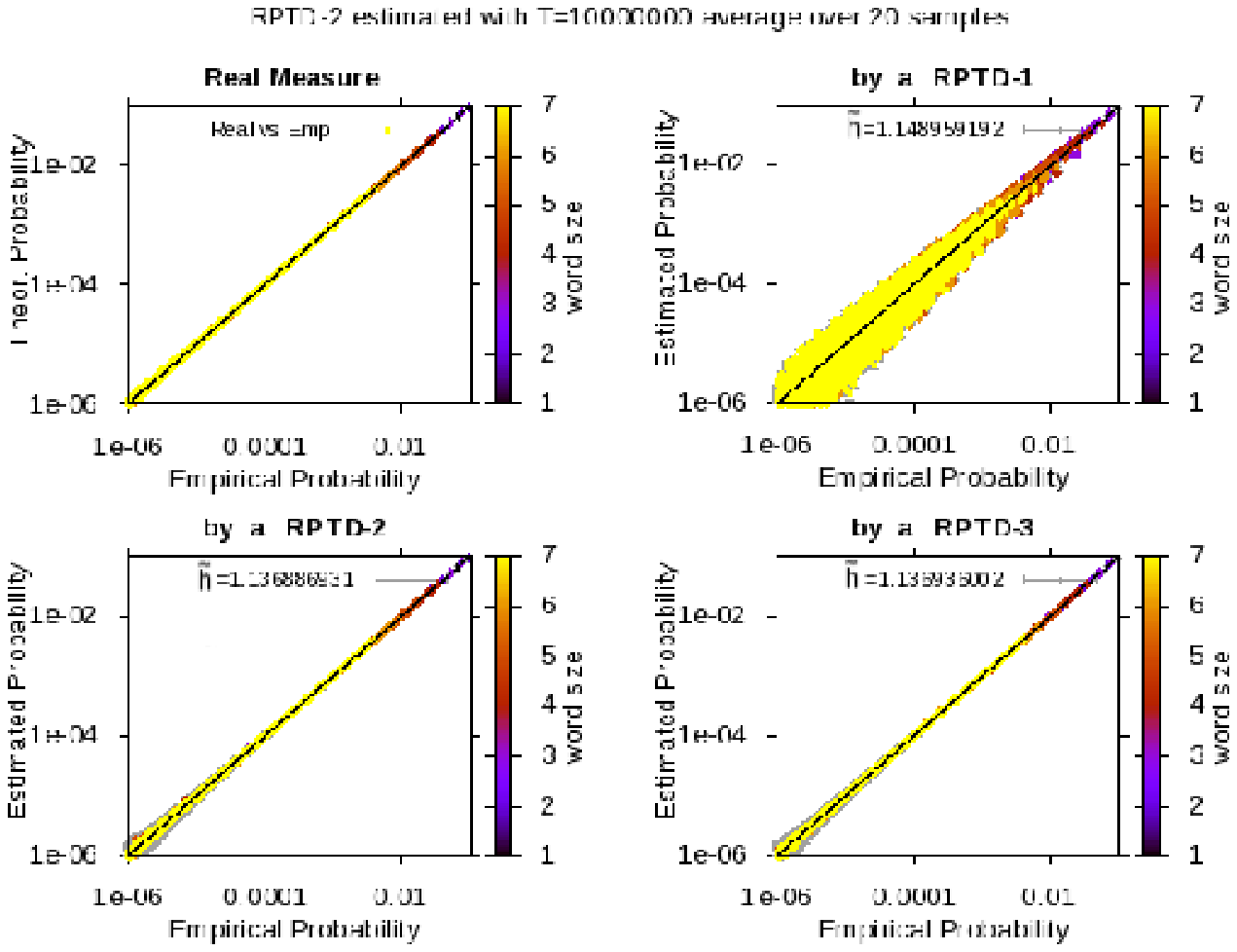}
\vspace{0.5cm}
\includegraphics[height=8.0cm, width=8.5cm]{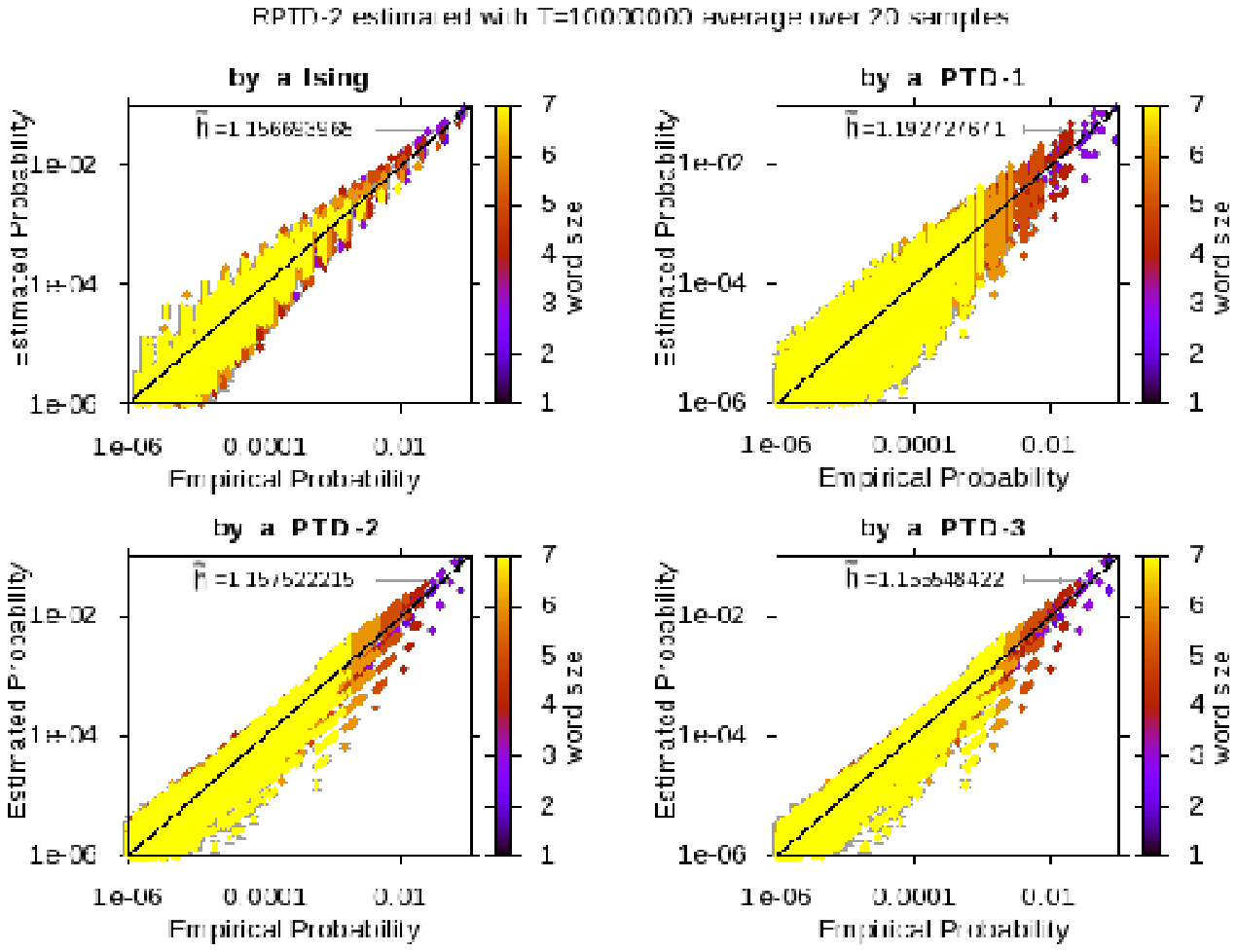}
\caption{
\footnotesize{
Figure 1 (top left) Expected probability $\mu_{\bpsi}$ versus  empirical probability $\pi_{\omega}^{(T)}(w)$;
Figure 2 (top right) to 8 (bottom right) Predicted probability versus empirical probability $\pi_{\omega}^{(T)}(w)$ 
for several models.  The generating potential is a \textbf{RPTD-2}.
}
}
\end{center}
\end{figure}

\begin{figure}\label{test2xb}
\begin{center}
\includegraphics[height=8.5cm, width=8.5cm]{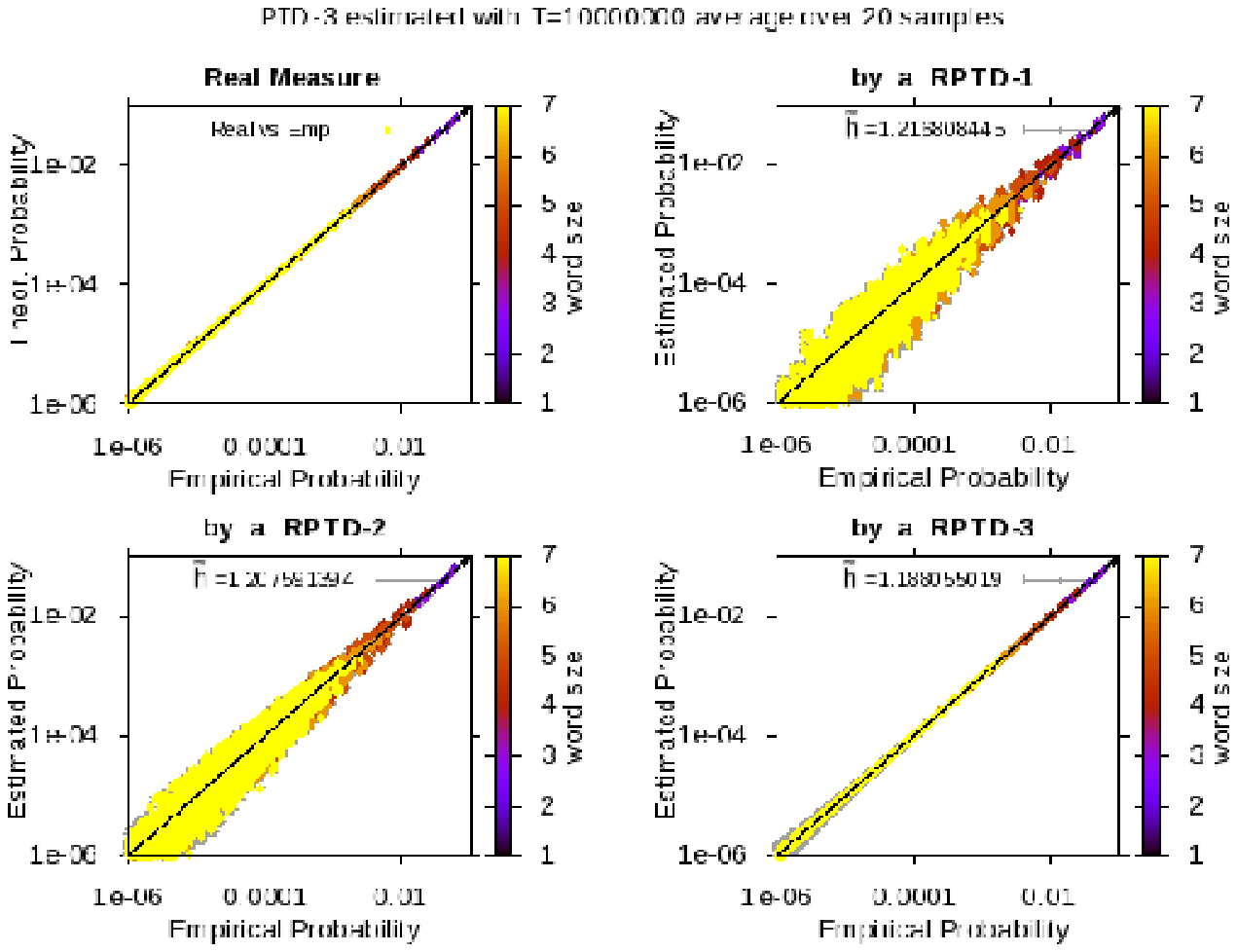}
\vspace{0.5cm}
\includegraphics[height=8.5cm, width=8.5cm]{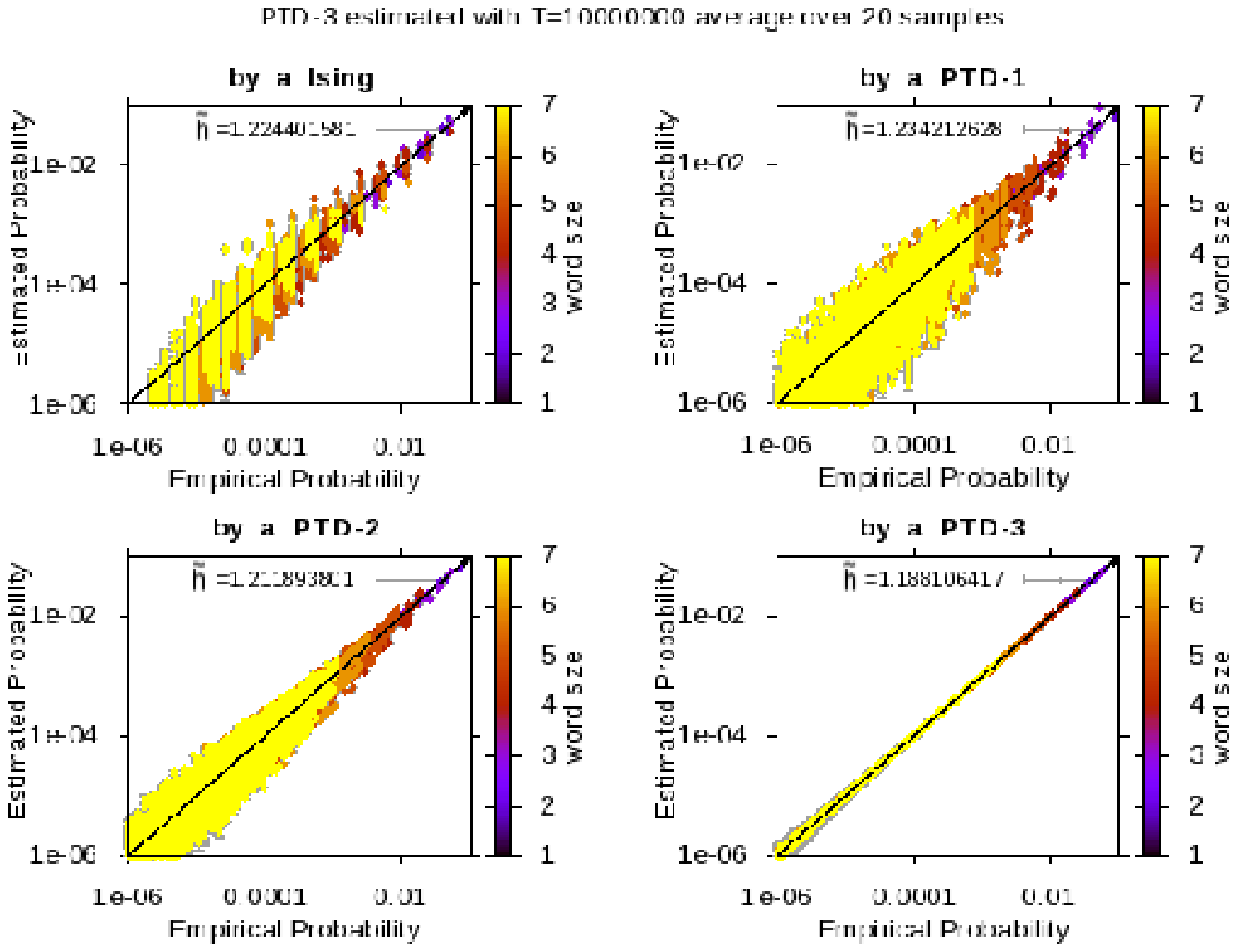}
\caption{
\footnotesize{Same as previous figure where generating potential is a \textbf{PTD-3}.}
}

\end{center}
\end{figure}



\bigskip

In order to extend the model comparison we introduce the following notations: let $w$ be a word (encoding a spiking pattern) of length $R$,   $P_{est}(w)$ its mean probability over trials  calculated with the estimated potential, $ P_{emp}(w)$ its mean empirical average over trials 
(i.e average of form (\ref{empav}) including a time average $\pTo$ and a sample average, where the samples
are contiguous pieces of the raster of length $T_0 \ll T$),
and $\sigma_{emp}(w)$ the standard deviation of $P_{emp}(w)$. We now describe the comparison methods. 

\begin{figure}\label{test2b}
 \includegraphics[height=6.5cm, width=6.5cm]{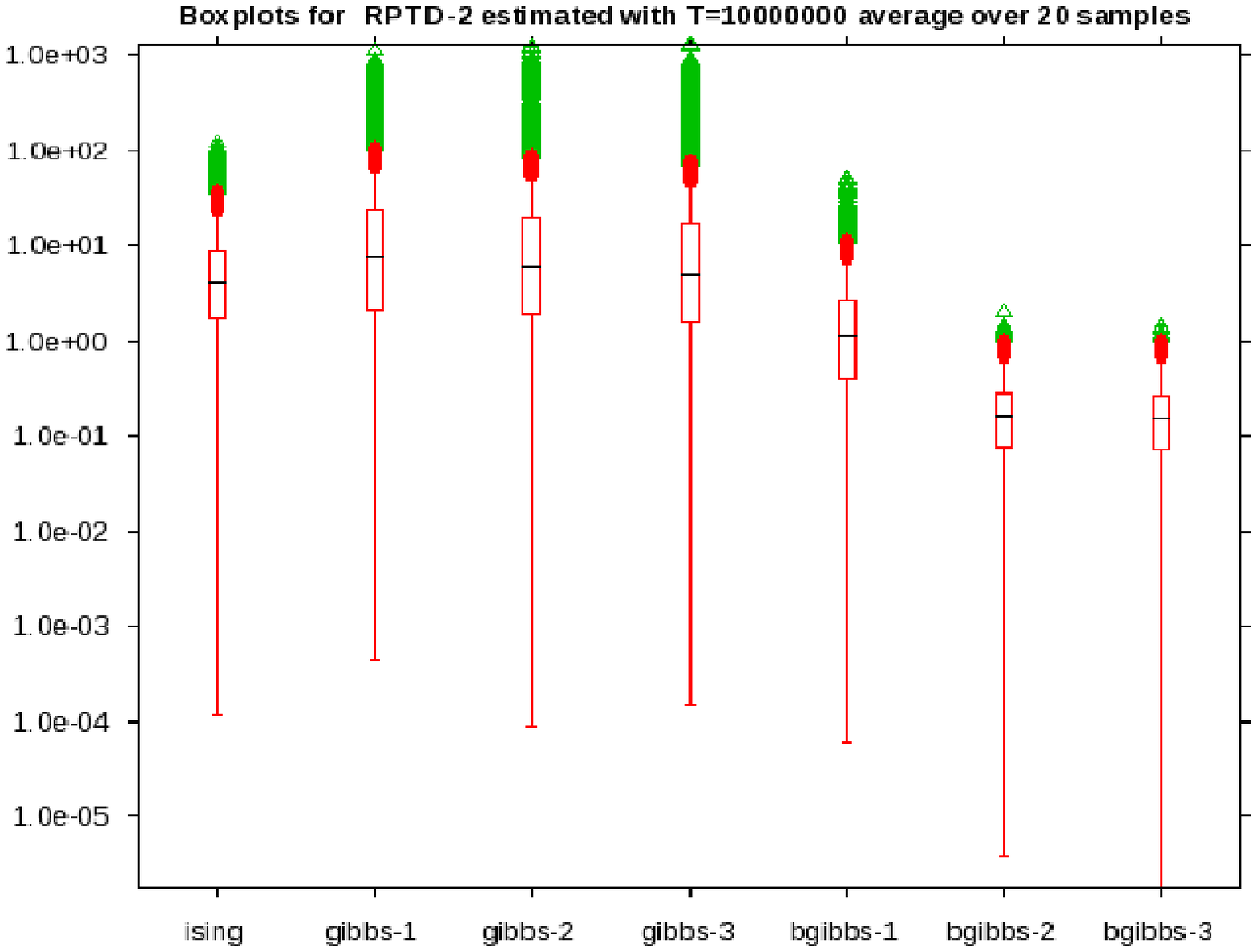}
\includegraphics[height=6.5cm, width=6.5cm]{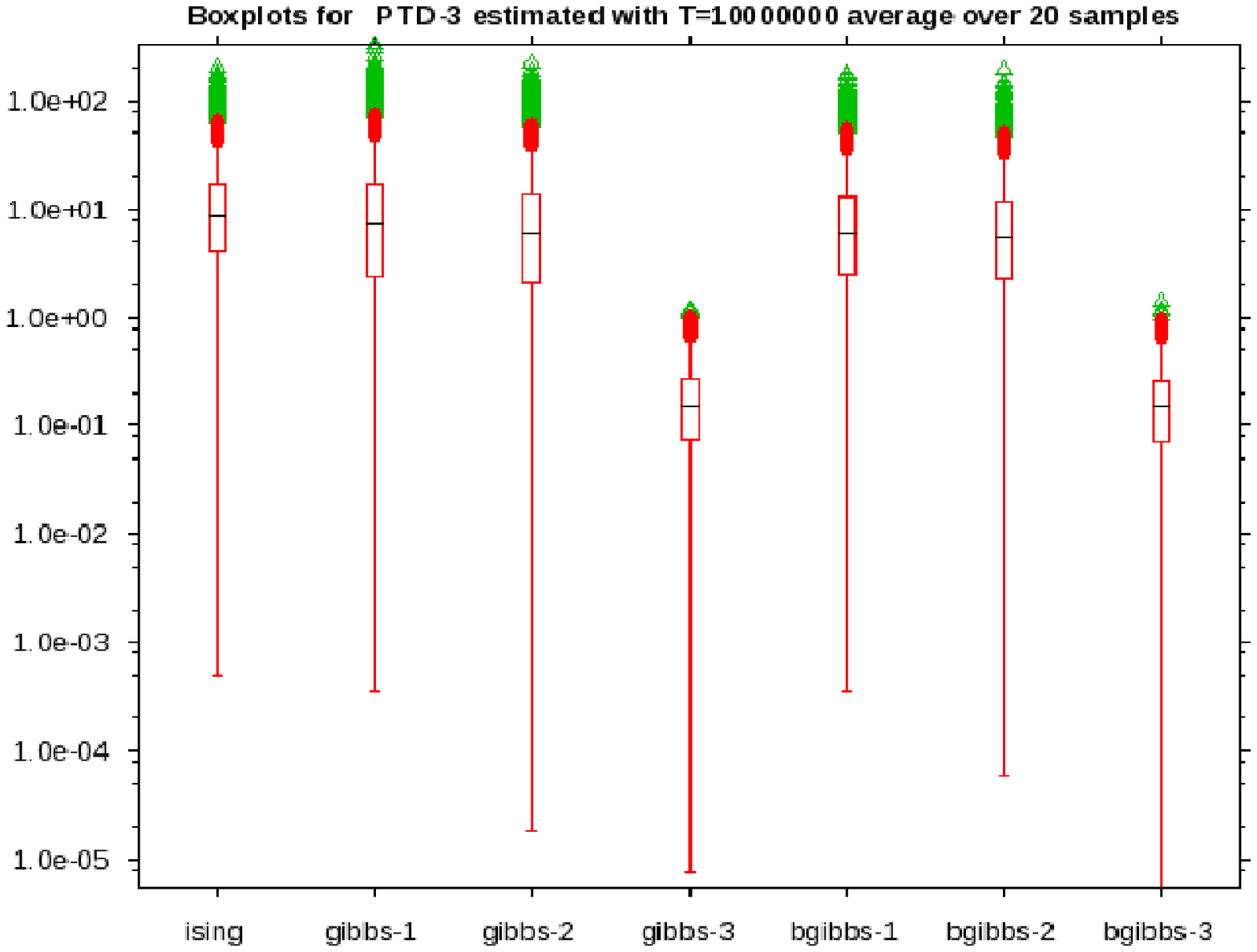}
\caption{\footnotesize{The box-plot (in log-scale) of the distributions of weighted deviations of word's probability versus their empirical probability, for several statistical models, using a generating potential of the form \textbf{(left)  RPTD-$2$}  and \textbf{(right) PTD-$3$}. Midliers Outliers (see footnote \ref{outliers}) are shown by red dots and extreme outliers by green dots.}}
\end{figure}

We first  use the box-plot method \cite{Frigge1989} which is intended to graphically depict groups of numerical data through their 'five-number summaries' namely: the smallest observation (sample minimum), lower quartile (Q1), median (Q2), upper quartile (Q3), and largest observation (sample maximum)\footnote
{
\label{outliers} The largest (smallest) observation is obtained using parameter dependent bounds, or ``fences'', to filter aberrant uninteresting deviations. Call $\beta=Q3-Q1$ and let $k$ denote the parameter value, usually between $1.0$ and $2.0$.
 Then the bound correspond to $Q3+k\beta$ for the largest observation (and for the smallest one to  $Q1-k\beta$).
 A point $x$ found above (below) is called ``mild-outlier'' if $ Q3+k< x <Q3+2k\beta $  (respectively, $ Q1-2k\beta< x <Q3-k\beta $) or extreme outlier if $ x > Q3+2k\beta $  (respectively, $ x < Q1-2k\beta$). 
We have used a fence coefficient $k=2.0$ to look for outliers.}.  
Figure \ref{test2b} shows, in log-scale, the box-plot for the distribution of the quantity defined as:
\beq \label{vareps}
\varepsilon(w)=|\left(P_{est}(w)-P_{emp}(w)\right)/\sigma_{emp}(w)|
\eeq
 that is taken as a weighted measure of the deviations. We have considered
this distribution when it takes  into
account, either
 all the words up to a given size $R_{max}$, or only the words of that given size.
There is no visual difference  for $R_{max}=7$. The results shows that only models containing 
the generating potential have the lower deviations value with very similar box.
 On the other hand a ``bad'' statistical model shows a much more extended error distribution .\\

Finally a  $\chi^2$ estimation is computed as $\chi^2= \frac{1}{N_{\text{words}}-L} \sum_w \varepsilon(w)^2 $
where $\varepsilon(w)$ is given by (\ref{vareps}). 
Values are reported in tables \ref{chitable}, using all words or only those of size $R_{max}$. 
Since the number of words is high, it is clear that the lower the error, the lower the $\chi^2$ estimated value. 
Note that $\chi^2$ test assumes Gaussian fluctuations about the mean value, which are satisfied
for finite-range Gibbs distributions, as can be easily seen by expanding the large deviations
function $I_l$ in (\ref{Conv_av_phil}) up to the second order in $\epsilon$. However,
when comparing two different Gibbs distributions it might be that the deviations
from the expected value of one Gibbs distribution compared to the expected
value of the other Gibbs distribution is well beyond the mean-square deviation of the Gaussian
fluctuations distribution,
giving rise to huge $\chi^2$ coefficients, as we see in the tables \ref{chitable}.  

\begin{table}
\caption{{\footnotesize {\bf $\chi^2$ coefficient calculated}: (left) with all words of size
$<7$;  (right) with  words of size $7$  only. See text for details.}}
{\small
\begin{tabular}{|l|c|c|}
\hline\noalign{\smallskip}
  \scriptsize{Estimating \textbackslash Generating}  & RPTD-2 & PTD-3 \\ 
\noalign{\smallskip}\hline\noalign{\smallskip}
 Ising& 135.427 &  415.965\\ 
PTD-1& 3146.17 &  564.396\\  
PTD-2&  3319.75& 290.93  \\ 
PTD-3& 2533.35 &  0.0571905 \\ 
RPTD-1& 13.9287 & 274.773  \\ 
RPTD-2& 0.0607027 &  223.516  \\
RPTD-3& 0.0556114&  0.0539691\\ 
\hline\noalign{\smallskip}
\end{tabular} 
\begin{tabular}{|l|c|c|}
\hline\noalign{\smallskip}
   \scriptsize{Estimating \textbackslash Generating} & RPTD-2 &PTD-3 \\ 
 \noalign{\smallskip}\hline\noalign{\smallskip}
 Ising& 121.825  &  347.502 \\ 
PTD-1& 2839.36 & 468.763  \\ 
PTD-2& 2537.39 & 229.255  \\ 
PTD-3& 2053.72 & 0.057065  \\ 
RPTD-1& 11.6167 &  218.458  \\ 
RPTD-2& 0.0605959 &  176.598 \\ 
RPTD-3& 0.0553242 & 0.0541206\\ 
\hline\noalign{\smallskip} 
\end{tabular}
}
\label{chitable}
\end{table}

\subsection{Spike train statistics in a simulated Neural Network }

Here we simulate an Integrate-and-Fire neural network whose spike train statistics is explicitely and rigorously
known \cite{cessac:10} while effects of synaptic plasticity on statistics have been studied in \cite{cessac-rostro-etal:09}.
 
\sssu{Network dynamics.}

The model is defined as follows. 
Denote  by $V_i$ the membrane potential of neuron $i$ and  $W_{ij}$ the synaptic weight of neuron $j$ over neuron $i$, $\Iei$ an external input on neuron $i$. Each neuron is submitted to noise, modeled by an additional input, $\sigma_B B_i(t)$,
with $\sigma_B >0$ and where the $B_i(t)$'s are Gaussian, independent, centered random wariable with variance $1$.
The network dynamics is given by:
\beq\label{FiBMS}
V_i(t+1)=\gamma \, V_i \left(1 - Z[V_i(t)] \right)+ \sum_{j=1}^N W_{ij}Z[V_j(t)]+ \Iei + \sigma_B B_i(t); \qquad i=1 \dots N,
\eeq
where $\gamma \in [0,1[$ is the leak in this discrete time model ($\gamma=1-\frac{dt}{\tau}$).
Finally, the function $Z(x)$ mimics a spike: $Z(x)=1$ if $x \geq \theta=1$ and $0$ otherwise, where
$\theta$ is the firing threshold.
As a consequence, equation (\ref{FiBMS}) implements both the integrate and firing regime.
It turns out that this  time-discretisation of the standard integrate-and-Fire neuron model, which as discussed  in e.g. \cite{izhikevich:03}, provides a rough but realistic approximation of biological neurons behaviors. 
Its dynamics has been fully characterized for $\sigma_B=0$ in \cite{cessac:08} while
the dynamics with noise is investigated in \cite{cessac:10}. Its links to more elaborated models
closer to biology is discussed in \cite{cessac-vieville:08}.

\sssu{Exact spike trains statistics.}

For $\sigma_B>0$ there is a unique Gibbs distribution in this model, whose potential
is explicitely known. It is given by:
\beq\label{phiBMS}
\phi(\seq{\omega}{-\infty}{0})=
\sum_{i=1}^N 
\left[\omega_{i}(0) 
\log\left(\pi\left(\frac{\theta-C_i(\uom)}
{\sigma_i(\uom)}\right)\right)+
\left(1-\omega_{i}(0)\right)
\log
\left(1-\pi\left(\frac{\theta-C_i(\uom)}{\sigma_i(\uom)}\right)\right)
\right],
\eeq
where $\pi(x)=\frac{1}{\sqrt{2\pi}}\int_x^{+\infty} e^{-\frac{u^2}{2}}du$,
$\uom=\seq{\omega}{-\infty}{-1}$,
$C_i(\uom)=
\sum_{j=1}^N W_{ij}x_{ij}(\uom)+\Iei \frac{1-\gamma^{t+1-\tau_i({\uom})}}{1-\gamma}$,
$x_{ij}(\uom)=\sum_{l=\tau_i(\uom)}^{t} \gamma^{t-l}\omega_j(l)$, 
$\sigma^2_i(\uom)=\sigma_B^2\frac{1-\gamma^{2(t+1-\tau_i(\uom))}}{1-\gamma^2}$.
Finally, $\tau_i(\uom)$ is the last time, before $t=-1$, where neuron $i$ has fired,
in the sequence $\uom$ (with the convention that $\tau_i(\uom)=-\infty$ for the
sequences such that $\omega_i(n)=0, \forall n < 0$). This potential  has infinite range
but range $R \geq 1$ approximations exist, that consist of replacing $\uom=\seq{\omega}{-\infty}{-1}$
by $\seq{\omega}{-R}{-1}$ in (\ref{phiBMS}). The KL divergence between the Gibbs measure of
the approximated potential and the exact measure decays like $\gamma^{R}$.
Finite range potentials admit a polynomial expansion of form (\ref{potential_expansion}). 

\sssu{Numerical estimation of spike train statistics}
Here we have considered only one example of model (\ref{FiBMS}) (more extended simulations
and results will be provided elsewhere). It consists of $4$ neurons, with a \textit{sparse}
connectivity matrix so that there are neurons without synaptic interactions. 
The synaptic weigths matrix is:

$$
\cW=\scriptsize{
\left(
\baR{ccccc}
0 & -0.568 &  1.77 &     0\\ 
1.6  &    0 & -0.174  &    0\\ 
0 & 0.332 & 0 &  -0.351\\ 
0  & 1.41  & -0.0602    &  0 \\
\eaR
\right)},
$$ 
while $\gamma=0.1, \sigma_B=0.25,\Iei=0.5$.

First, one can compute directly the theoretical entropy of the model using the results
exposed in the previous section: the entropy of the range-$R$ approximation, that can be computed
with our formalism, converges exponentially fast with $R$ to the entropy of the infinite range potential.
For these parameters, the asymptotic value is $h=0.57$.\\

Then, we generate a raster of length $T=10^7$ for the $4$ neurons and we compute the KL divergence
between the empirical measure and several potentials including: 

\bit
\item{(i)} The range-$R$ approximation of (\ref{phiBMS}), denoted $\phi^{(R)}$. Note that
$\phi^{(R)}$  does not contain all monomials. In particular, \textit{it does not have the Ising term
(the corresponding coefficient is zero)}.
\item{(ii)} A Bernoulli model $\phi^{Ber}$; 
\item{(iii)} An Ising model $\phi^{Is}$; 
\item{(iv)} A one-time step Ising Markov model (as proposed in \cite{marre-boustani-etal:09})
$\phi^{MEDF}$ \footnote{or equivalently, a \textbf{RPTD-$1$} from  (\ref{pot2})}  ;
\item{(v)} A range-$R$ model containing all
monomials $\phi^{all}$.
\eit

Here we can compute the KL divergence since we known the theoretical entropy.
The results are presented in the table (\ref{TBMS}). Note that the estimated
 KL divergence of range-$1$ potentials slightly depend on $R$ since the RPF operator,
and thus the pressure, depend on $R$.

\begin{table}[h]
\caption{\footnotesize{\label{TBMS} Kullback-Leibler divergence between the empirical measure
of a raster generated by (\ref{FiBMS}) (See text for the parameters value) and the Gibbs
distribution, for several statistical models.}}
{\small
\begin{center}
\begin{tabular}{cc}
\begin{tabular}{l|cccccc}
\hline\noalign{\smallskip}
&$\phi^{(R)}$&$\phi^{Ber}$&$\phi^{Is}$&$\phi^{MEDF}$&$\phi^{all}$\\ 
\noalign{\smallskip}\hline\noalign{\smallskip}
 R=1 & 0.379  &0.379 & 0.312 & 1.211 & 0.309  \\
 R=2 & 0.00883  & 0.299871 & 0.256671 & 0.257068 & 0.0075 \\
 R=3 & -0.001  & 0.250736 & 0.215422 & 0.200534 & 0.0001  \\
\hline\noalign{\smallskip}
\end{tabular} 
\end{tabular} 
\end{center}
}
\end{table}

We observe that our procedure recovers the fact that the range-$R$ potential $\phi^{(R)}$
is the best to approximate the empirical measure, in the sense that it minimizes the KL
divergence and that it has the minimal number of terms ($\phi^{all}$ does as good as
$\phi^{(R)}$ for the KL divergence but it contains more monomials whose coefficient (almost) vanish in the estimation).

\sssu{Synaptic plasticity.}

Here the neural network with dynamics given by ~(\ref{FiBMS}) has been submitted to the STDP rule ~(\ref{Rexample}). The goal is to check the validity of the statistical model given by ~(\ref{STDP}), predicted in \cite{cessac-rostro-etal:09}. We use spike-trains of length $T=10^7$ from a simulated  network with $N=10$ neurons.

Previous numerical explorations of the noiseless case, $\sigma_B=0$, have shown \cite{cessac:08,cessac-vieville:08} that  a network of $N$ such neurons,
 with fully connected graph, where  synapses are taken randomly from a distribution ${\cal N}(0,\frac{C^2}{N})$, where $C$ is a control parameter,
 exhibits generically a dynamics with very large periods in determined regions of the parameters-space $(\gamma,C)$. 
On this basis, we choose; $N=10,\; \gamma=0.995,\; C=0.2$. 
The external current $\Ie$ in eq. (\ref{FiBMS}) is given by
 $\Iei=0.01$ while $\sigma_B=0.01$.
 Note that fixing a sufficiently large average value for this current avoids a situation where neurons stops firing after a certain time (``neural death'').

\begin{figure}\label{test3}
\begin{center}
 \includegraphics[height=8.5cm, width=8.5cm]{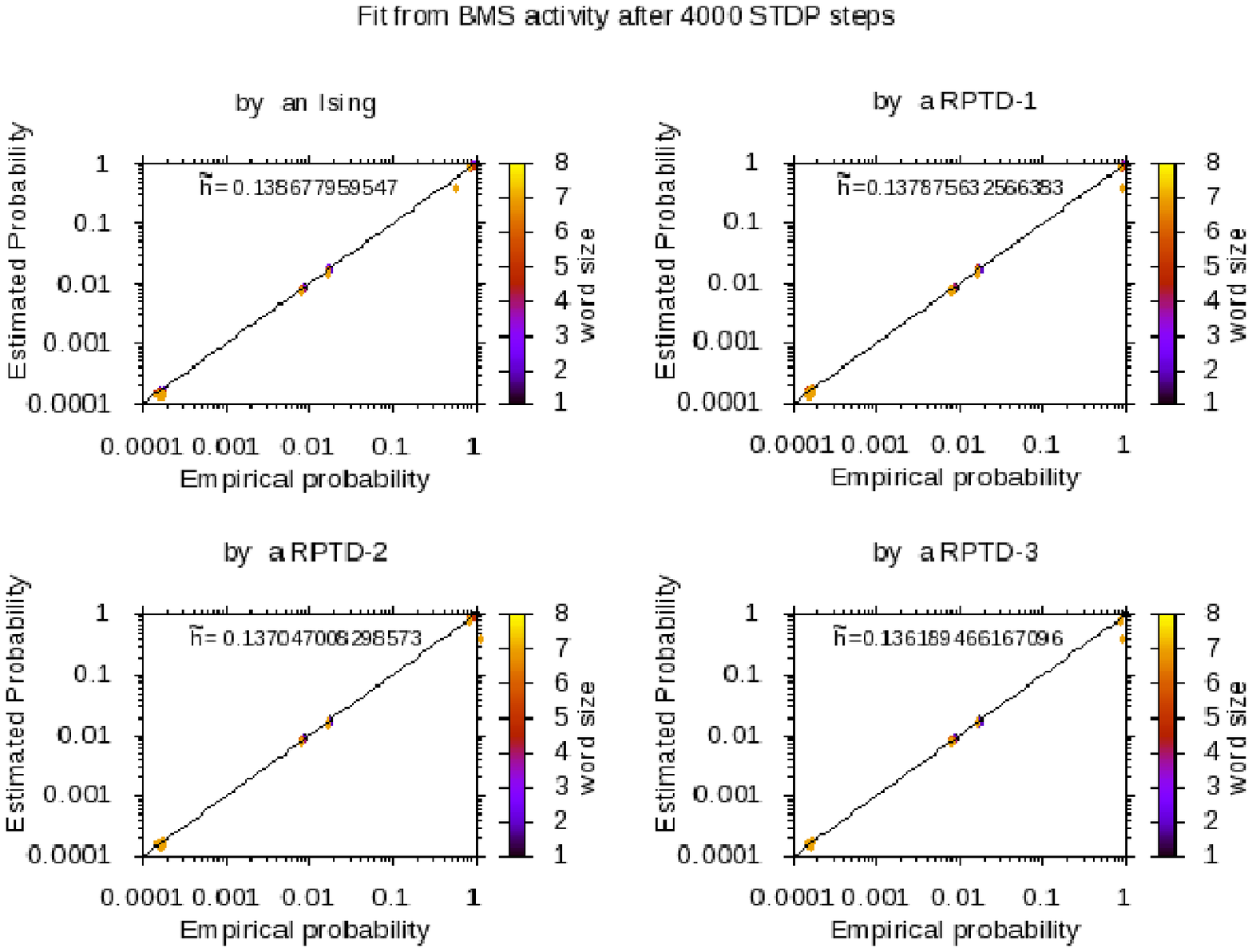}
\vspace{0.5cm}
\includegraphics[height=8.5cm, width=8.5cm]{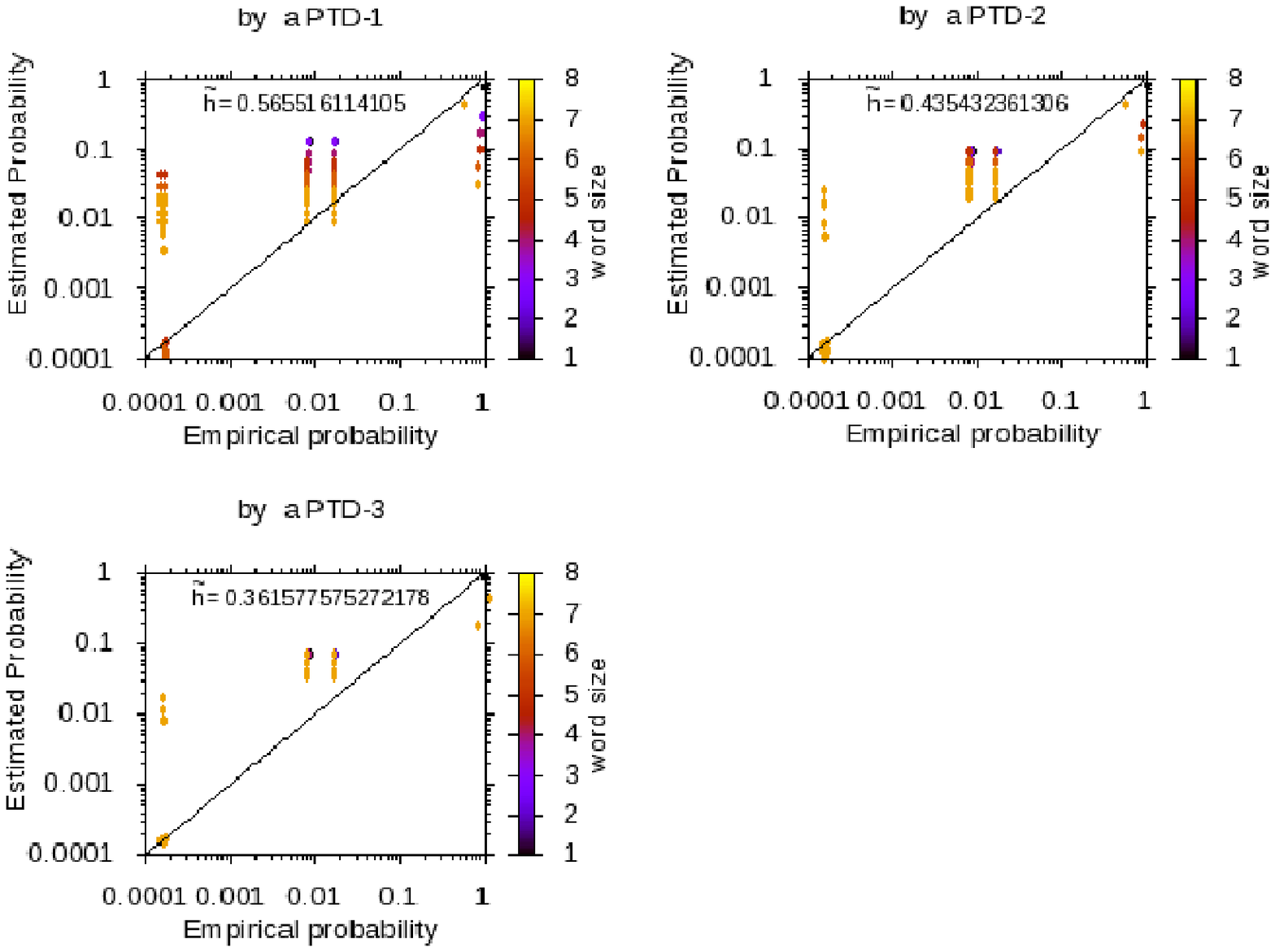}
\caption{\footnotesize{The probability of  words of different sizes 
 predicted by several statistical models from ~(\ref{pot2}) versus empirical probability $\pi_{\omega}^{(T)}(w)$
obtained from a spike train  generated by dynamics (\ref{FiBMS}) after 4000 epochs of adaptation.The $\tilde{h}$ value ~(\ref{criterion}) for each fitting model is shown inside the graphic. The potential is a  pair potential of the form ~(\ref{STDP2}). Recall that \textbf{RPTD} Models include firing rates but \textbf{PTD} models do not.}}
\end{center}
\end{figure}

We register the activity after $4000$ steps of adaptation with the STPD rule proposed in ~(\ref{Rexample}). In this context we expect the potential for the whole population to be of the form ~(\ref{STDP}) and for a subset of the population of the form ~(\ref{STDP2}). Therefore, we choose randomly 2 neurons among the $N$ and we construct from them the prefix-tree. Then, for the 2 neuron potentials forms from ~(\ref{pot2}),  we estimate the  coefficients that minimizes the Kullback-Leibler divergence.The probability of  words of different sizes predicted by several statistical models from ~(\ref{pot2}) versus empirical probability $\pi_{\omega}^{(T)}(w)$ obtained from a spike train and the corresponding $\tilde{h}$ value of the estimation process for a fixed pair of neurons are shown on figure (\ref{test3}). \\
Results depicted on figure (\ref{test3}) show, on one hand, that the statistics is well fitted by  ~(\ref{STDP2}). 
Moreover, the best statistical models, are those including rate terms (the differences between their KL value is two orders of magnitude smaller that within those not disposing of rate terms). 
We also note  that for the words with the smallest probability values, the potential do not yields a perfect 
matching due to finite size effects (see fig (\ref{test3})). 
Especially, the small number of events due to low firing rates of neurons makes more sensitive the relation 
between the length of observed sequences (word size)  and the spike-train length necessary to provide a good sampling and hence a reliable empirical probability.

\subsubsection{ Additional tests: the non-stationary case}\label{SNonStat}
Here we present results of the parameter estimation method applied to a spike train with statistics governed by a non-stationary statistical model of range $1$, i.e. with time varying coefficients for rate or synchronization terms. Since the generation of spike-trains corresponding to more general higher time-order non-stationary process is not trivial, these potentials with higher range values will be analyzed in a forthcoming paper.

In the following we use an Ising potential form ~(\ref{pot2}) with time-varying coefficients $\bpsi = (\omega) =\lambda_1(t)\,\omega_1(0)+\lambda_2(t)\,\omega_2(0)+ \lambda_3(t)\,\omega_1(0)\,\omega_2(0).$. The procedure to generate a non stationary spike-train of length $T$ is the following.
We fix a time dependent form for the  3 coefficients $\lambda_i(t)$.
From the initial value of the $\lambda_i's$ (say at time $t$) we compute the invariant measure
of the RPF operator. From this, we draw a 
 Chapman-Kolmogorov equation ~(\ref{ChapKol}) with a time dependent RPF operator computed using the
next coefficient values $\lambda_i(t+1)$.

With the generated spike-train, we perform the parameter estimation, but computing the empirical average over an small fraction of it which means a time window of size $T_0=\frac{T}{M} <<T$. Then, we slide the observation window and estimate again the coefficients value. We have verified that estimation procedure can recover correctly 
the coefficient values, for several types of time dependence, provided their variations be not too fast,
 and that the sliding window size be not too large with respect to $T$.
 We present the reconstruction of the parameters with a sinusoidal time-dependence given by $\lambda_0(t)=0.4+0.3\sin\big(\frac{4\pi t}{T-T_0}\big)$.

\begin{figure}[!ht]
\begin{center}
\includegraphics[height=6.2cm, width=6.0cm]{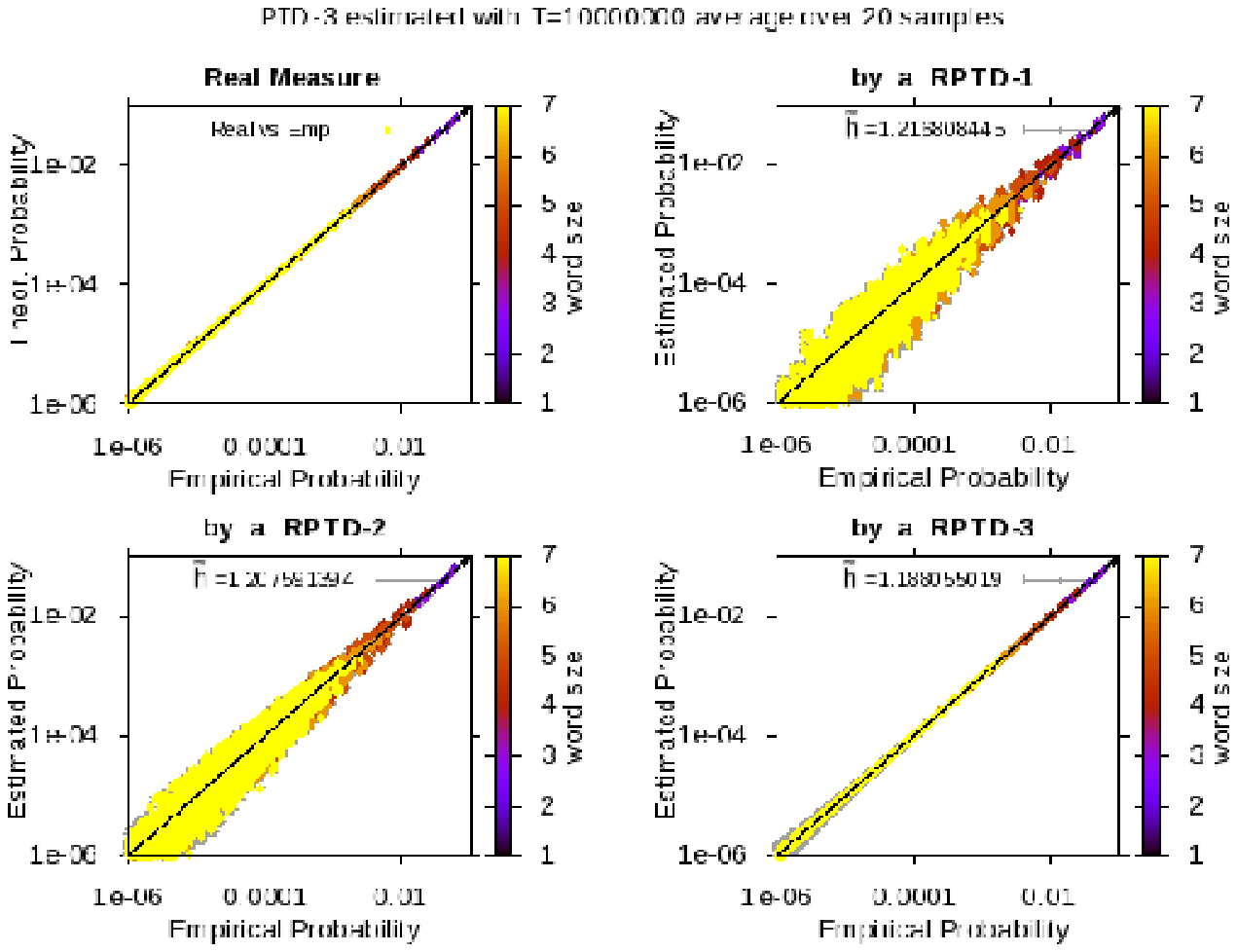}
\includegraphics[height=6.2cm, width=6.0cm]{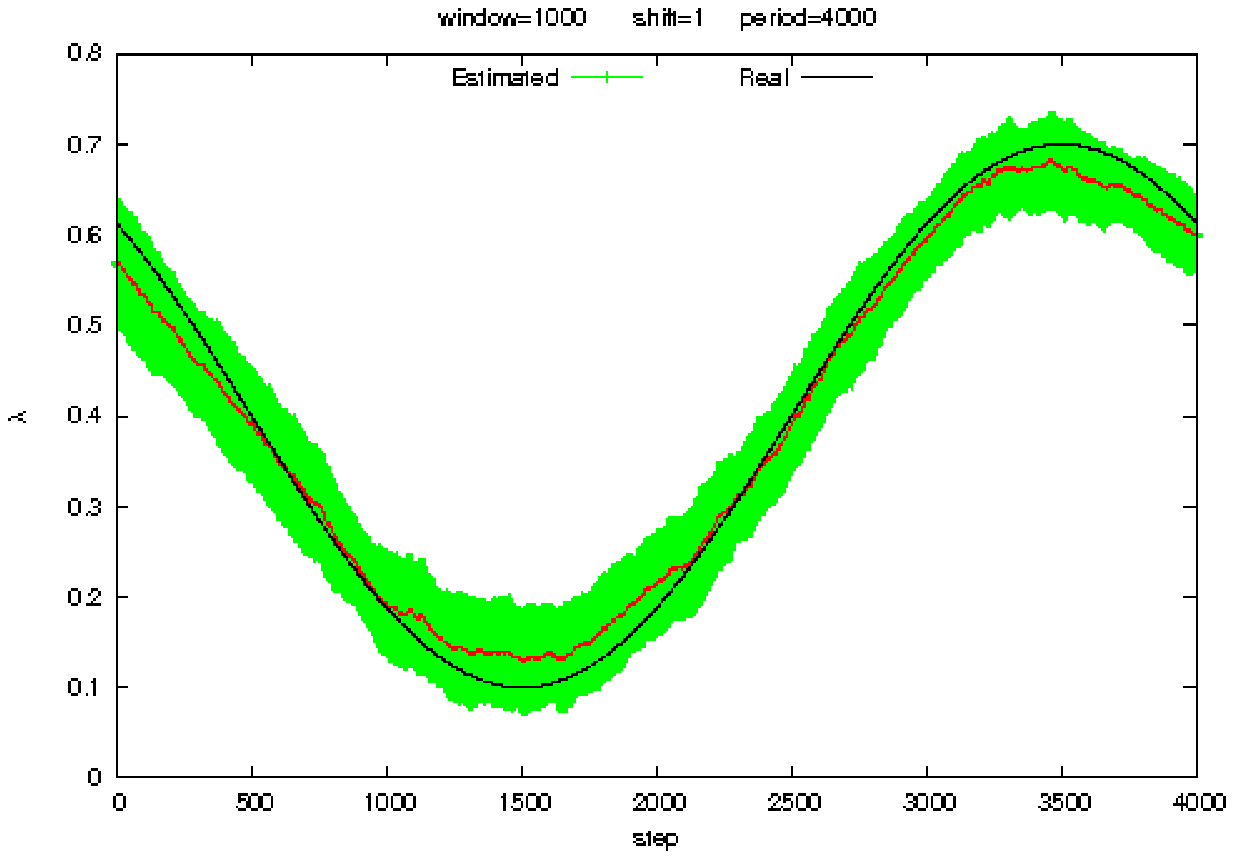}
\end{center}
\caption{\footnotesize{{\bf Estimation of coefficients on a Non-Stationary process genated by an Ising model and sinousoidal time dependence}. Real value(black) and estimated parameter with its error bars (green) computed over 20 trials. The time shift is $\tau=1$ , Window size is fixed $1000$, but oscillation period corresponds to $2000$ (left) and $4000$ (right). }}
\label{testns}
\end{figure}

\su{Discussion and conclusion}

\ssu{Comparison with existing methods}

Let us first summarize the advantages and drawbacks of our method compared with
the existing ones. For this, we list some keywords in the approaches used by the community
and discuss the links with our own work.

\bit
\item\textbf{Maximum entropy.}
The formalism that we use corresponds to a maximum entropy method but without limitations
on the number or type of contraints. Actually, on mathematical grounds,
it allows infinitely many constraints. Moreover, we do not need to compute the entropy.

\item\textbf{Markovian approaches.}
Our method is based on a Markovian approach where the memory depth of the Markov chain can be
arbitrary long (actually the formalism that we use allows to theoretically consider  processes with infinite memory, called
\textit{chains with complete connections \cite{maillard:07}}, see \cite{cessac:10} for an application to spike train statistics).
As we developed, the link between the potential extracted from the maximum entropy principle,
by fixing \textit{ad hoc} observables, and a Markov chain is not straightforward, since a potential
of this kind is not normalized.

\item\textbf{Monte-Carlo methods.}
Equation (\ref{ChapKolNN}) allows us to generate spike trains Gibbs-distributed
with and arbitrary potential (non normalized). The convergence is ensured 
by eq. (\ref{ConvVect}). We emphasize that we do not need to assume detailed balance. 
Instead, we impose a technical assumption (primitivity of the Ruelle-Perron-Frobenius matrix)
which is more general than detailed balance. On the opposite, if this assumption does not
hold then the unicity of the Gibbs distribution is not guarantee and, in this case, the determination
of spike train statistics from empirical data becomes even more cumbersome.

\item\textbf{Determining an effective synaptic connectivity between neurons.}
Interactions between neurons occur via synapses (or gap junction). This interaction is
not instantaneous, it requires some delay. As a matter of fact, estimating the synaptic conductances
via the spike statistics requires therefore to consider time-dependent potentials.
Our formalism allows this. Determining an effective synaptic connectivity between neurons from spike trains
will be the subject of a forthcoming paper. 

\item\textbf{Boltzmann learning.}
Our approach can be viewed as ``Boltzmann learning'' (as presented e.g. in \cite{roudi-tyrcha-etal:09}) without restrictions
on the parameters that we learn, without using a Monte Carlo approach (which assumes detailed balance), and uses
a criterion which is strictly convex.

\item\textbf{Performances.} At its current implementation level, the proposed method allows us to analyze the statistics of small groups (up to 8/12) of neurons.
The parametric statistical potential of Markov processes up to range 16/20 is calculable, thus considering up to 2$^{20}$ states for the process. The implementation considers several well-established numerical methods, in order to be applicable to a large set of possible data. With respect to the state of the art, this method allows us to consider
non-trivial statistics (e.g. beyond rate models and even models with correlation),
thus targeting models with complex spike patterns. This method is in a sense the next step after Ising models, 
known as being able to represent a large but limited part of the encoded information (e.g. \cite{schneidman-etal:06,Mezard2009}).
Another very important difference with respect to other current methods is that we perform the explicit variational optimization of a well defined quantity, 
i.e., the KL-divergence between the observed and estimated distributions. The method proposed here does not rely on Monte Carlo Markov Chain methods but on a spectral computation based on the RPF operator, providing exact formula, while the spectral characteristics are easily  obtained from standard numerical methods. 

The main drawback of our method
is that it \textit{does not allow to treat a large number of neurons and simultaneously a large range}.
This is due to the evident fact that the number of monomials combinatorically increases
as $N$, $R$ growth. However, this is not a problem intrinsic to our approach but to parametric
estimations potentials of the form (\ref{psi}). We believe that other form of potential
could be more efficient (see \cite{cessac:10} for an example).
We also want to emphasize that, when considering Ising like statistics our algorithm is \textit{less performant} than the existing ones (although improvements
in speed and memory capacity thanks to the use of parallel computation algorithms remain an open and natural developpement path), for the simple reason that the latter has been developed and optimized using the tremendous results existing in statistical physics, for spins systems. Their extensions to models of the general form (\ref{psi})
seems rather delicate, as suggested by the nice work in \cite{marre-boustani-etal:09} where extension between
the $1$-step Markov case is already cumbersome.

\item\textbf{Mean-field methods.}
Mean-field methods aim at computing the average value of observables
(``order parameters'') relevant for the characterisation of statistical properties of the system.
Typical examples are magnetisation in ferromagnetic models (corresponding to
rates in spiking neurons models), but more elaborated order parameters are known
e.g. in spin glasses \cite{mezard:87} or in neural networks \cite{sompolinsky-etal:88}.
 Those quantities obey equations (usually called mean-field equations) which are, in most cases,
not explicitely solvable. Therefore, approximations are proposed from the simplest
(naive mean-field equations) to more complex estimations, with significant results developed in
the realm of spins systems (Ising model, Sherrington-Kirckpatrick spin glass model
\cite{sherrington-kirkpatrick:75}). Examples are the replica method \cite{mezard:87},
Thouless-Anderson-Palmer equations
\cite{thouless-anderson-etal:77}, the Plefka expansion \cite{plefka:82}, or more recently e.g. the Sessak-Monasson
approximation \cite{sessak-monasson:09} (for a recent review
on mean-field methods see \cite{opper-saad:01}). Since the seminal paper by Schneidman and collaborators  \cite{schneidman-berry-etal:06} they have also been applied to spike trains statistics analysis assuming that neurons dynamics
generates a spike statistics characterized by a Gibbs distribution with an Ising Hamiltonian.
In their most common form these methods do not consider dynamics (e.g time correlations) and
their extension to the time-dependent case (e.g. dynamic mean-field methods) is far from being 
straightforward (see e.g. \cite{sompolinsky-zippelius:82,sompolinsky-etal:88,benarous-guionnet:95,samuelides-cessac:07,faugeras-touboul-etal:08} for examples of such developments). Moreover, exact mean-field equations 
and their approximations usually
only provide a probability measure at positive distance to
 the true (stationary) probability measure of the system (this distance
can be quantified in the setting of information geometry using e.g. the KL distance \cite{amari:00}).
This is the case whenever the knowledge of the sought order parameters is not sufficient to determine
the underlying probability. 

The present work can, in some sense, be interpreted in the realm of mean-field approaches.
Indeed, we are seeking an hidden Gibbs measure and we have only information about
the average value of ad hoc observables. Thus, equation
(\ref{Gener}) is a mean-field equation since it provides the average value
of an observable with respect to the Gibbs distribution. There are therefore $L$
such equations, where $L$ is the number of monomials in the potential $\bpsi$.
Are all these equations relevant ? If not, which one are sufficient to determine univoquely
the Gibbs distribution ? Which are the order parameters ?  
The method consisting of providing a hierarchy of mean-field approximations which
starts with the Bernoulli model (all monomials but the rate terms are replaced by a constant),
then Ising (all monomials but rate and spatial correlations are replaced by a constant), while
progressively diminishing the KL divergence allows to answer the question of the relevant
order parameters and can be interpreted as well in the realm of information geometry.
This hierarchical approach is a strategy  to cope with the problem of combinatorial explosion
of terms in the potential when the number of neurons or range increases. But the form of potential
that we consider does not allow a straightforward application of the methods inherited from 
statistical mechanics of spin systems. As a consequence, we believe that instead
of focusing too much on these methods it should be useful to adopt
technics based on large deviations (which actually allows the rigorous fundation of dynamic mean field
methods for spin-glasses \cite{benarous-guionnet:95} and neural networks
\cite{samuelides-cessac:07,faugeras-touboul-etal:08}). 
This is what the present formalism offers.

\eit

\ssu{Conclusion and perspectives}

The thermodynamic formalism 
allows us to provide closed-form calculations of interesting parameters related to spectral properties of the RPF operator.
We, for instance, propose an indirect estimation of the entropy, via an explicit formula.
We also provide numbers for the average values of the related observable, probability measure, etc.. 
This means that as soon as we obtain the numerical values of the Gibbs distribution up to some numerical precision, 
all other statistical parameters come for free without additional approximations.

A step further, the non-trivial but very precious virtue of the method is that it allows us to efficiently compare  models. 
We thus not only estimate the optimal parameters of a model, but can also determine among a set of models which model is the most relevant. 
This means, for instance, that we can determine if either only rates, or rates and correlations matters, for a given piece of data.
Another example is to detect if a given spike pattern is significant, with respect to a model not taking this pattern into account.
The statistical significance mechanism provides numbers that are clearly different for models corresponding or not to a given empirical distribution,
providing also an absolute test about the estimation significance.
These elements push the state of the art regarding statistical analysis of spike train a step further.\\

At the present state of the art, the present method is limited by three bounds.

First of all, the formalism is developed for a stationary spike-train, i.e. for which the statistical parameters are constant. 
This is indeed a strong limitation, especially in order to analyze biological data, though several related approaches consider the same restrictive framework.
This drawback is overcome at two levels. At the implementation level we show here how using a sliding estimation window and assuming an adiabatic, i.e. slowly varying, 
distribution we still can perform some relevant estimation. In a nutshell, 
the method seems still usable and we are now currently investigating this on both simulated and biological data, this being another study on its own.
At a more theoretical level, we are revisiting the thermodynamic formalism developed here for time varying parameters 
(in a similar way as the so called inhomogeneous Poisson process with time varying rates). 
Though this yields non-trivial developments  beyond the scope of this work, it seems that we can generalize the present formalism in this direction.

Secondly, the present implementation has been optimized for dense statistical distributions, i.e., in the case where almost all possible spike combinations are observed.
Several mechanisms, such as look-up tables, make this implementation very fast. However, if the data is sparse, as it may be the case for biological, a dual implementation
has to be provided using data structure, such as associative tables, well adapted to the fact that only a small amount of possible spike combinations are observed.
This complementary implementation has been made available and validated against the present one. This is going to analyze sparse Markov processes up to range much higher
than 16/20. Again this is not a trivial subject and this aspect must be developed in a next study as well as the applicability of parallel computing alternatives ( e.g. sparse matrix storage, parallel fast-eigenvalue algorithms, etc.).

Finally, given an assembly of neurons, every statistical tools available today provide only the analysis of the statistics a small subset of neurons, and it is known
that this only partially reflects the behavior of the whole population \cite{latham-roth-etal:06}. 
The present method for instance, is difficult to generalize to more than 8/10 neurons because of the incompressible algorithmic complexity of the formalism although parallel computation techniques might be helpful. However, the barrier is not at the implementation level, but at the theoretical level, since effective statistical general models (beyond Ising models) allow for instance to analyze statistically large spiking patterns such as those observed 
in synfire chains \cite{hertz:97} or polychronism mechanisms \cite{paugam-moisy-etal:08}. 
This may be the limit of the present class of approaches, and things are to be thinked differently.
We believe that the framework of thermodynamic formalism and links to Statistical Physics is still a relevant source of methods for such challenging perspectives.

\subsection*{acknowledgements}
We are grateful to F. Grammont, C. Malot, F. Delarue and P. Reynaud-Bourret for helpful discussions and Adrien Palacios for his precious remarks and profound scientific questions at the origin of main aspects of the present work. Partially supported by the ANR MAPS \& the MACACC ARC projects and PhD.D-fellowship from Research Ministry to J.C Vasquez.
%
%

\bibliographystyle{plain}
{\scriptsize \bibliography{../../../../../tex/biblio-gibbsgramme,../../../../../tex/odyssee,../../../../../tex/biblio}}

\end{document}

%% file: macro.tex
\newcommand{\ed}{\end{document}}
\newcommand{\beq}{\begin{equation}}
\newcommand{\eeq}{\end{equation}}
\newcommand{\nid}{\noindent}
\newcommand{\ben}{\begin{enumerate}}
\newcommand{\een}{\end{enumerate}}
\newcommand{\bit}{\begin{itemize}}
\newcommand{\eit}{\end{itemize}}
\newcommand{\baR}{\begin{array}}
\newcommand{\eaR}{\end{array}}
\newcommand{\su}[1]{\section{#1}}
\newcommand{\ssu}[1]{\subsection{#1}}
\newcommand{\sssu}[1]{\subsubsection{#1}}
\newcommand{\deq}{\stackrel {\rm def}{=}}
\newcommand{\bea}{\begin{eqnarray}}
\newcommand{\eea}{\end{eqnarray}}

\newcommand\D{\displaystyle}

\newtheorem{definition}{Definition}
\newcommand{\bdf}{\begin{definition}}
\newcommand{\edf}{\end{definition}}

\newtheorem{theorem}{Theorem}
\newcommand{\bth}{\begin{theorem}}
\newcommand{\enth}{\end{theorem}}

\newtheorem{proposition}{Proposition}
\newcommand{\bp}{\begin{proposition}}
\newcommand{\ep}{\end{proposition}}
\newtheorem{corrollary}{Corrollary}
\newcommand{\bcor}{\begin{corrollary}}
\newcommand{\ecor}{\end{corrollary}}
\newtheorem{lemma}{Lemma}
\newcommand{\blem}{\begin{lemma}}
\newcommand{\elem}{\end{lemma}}
\newtheorem{conjecture}{Conjecture}
\newcommand{\bconj}{\begin{conjecture}}
\newcommand{\econj}{\end{conjecture}}

\newcommand\bloc[2]{\left[\omega\right]_{#1}^{#2}}
\newcommand\cyl[3]{\left[{#1}\right]_{#2}^{#3}}
\newcommand\seq[3]{{#1}_{#2}^{#3}}

\newcommand{\setZ}{\mathbbm{Z}}

\newcommand{\setR}{\mathbbm{R}}
\newcommand\bbbr{{\sf I\!R}}

\newcommand\Vm{V_{min}}
\newcommand\VM{V_{max}}

\newcommand\X{{\bf X}}
\newcommand\Q{{\bf Q}}

\newcommand\Wij{W_{ij}}

\newcommand\ld{r_{d}}

\newcommand\cP{{\cal P}}

\newcommand\cT{{\cal T}}

\newcommand\pTo{{\pi_{\tom}^{(T)}}}

\newcommand\mpg{{\mu_{\bspsi}}}

\newcommand\bl{\boldsymbol\lambda}
\newcommand\bphi{\boldsymbol\phi}

\newcommand\bpsi{\boldsymbol\psi}
\newcommand\bPsi{\boldsymbol\Psi}
\newcommand\bom{\boldsymbol\omega}

\newcommand\bv{{\bf v}}

\newcommand\bspsi{\scriptstyle{\boldsymbol\psi}}

\newcommand\tom{\omega}

\newcommand\sit{{\sigma^t}}

\newcommand\Spn{\Sigma^{(n)}}

\newcommand\pres{P(\bpsi)}

\newcommand\omei{\omega_i}
\newcommand\omej{\omega_j}

\newcommand\Ie{{\bf{I}}^{(ext)}}
\newcommand\Iei{I^{ext}_i}

\newcommand\cH{{\cal H}}
\newcommand\cI{{\cal I}}
\newcommand\cN{{\cal N}}
\newcommand\cM{{\cal M}}

\newcommand\cW{{\cal W}}

\newcommand\Con{\left[\tom\right]_{0,n-1}}